\documentclass[11pt]{article}
\usepackage{geometry}              
\usepackage[utf8]{inputenc}
\usepackage{graphicx}
\usepackage[left]{lineno}
\usepackage{float}
\usepackage{tikz}

\newcommand{\epem}{$e^{+} e^{-}$}
\newcommand{\belleii}{Belle~II\,}
\newcommand{\ztrig}{$z$-Trigger\,}
\newcommand{\cmss}{\rm{cm}$^{-2}$\rm{s}$^{-1}$} 
\newcommand{\drift}{drift~time\,}
\newcommand{\drifts}{drift~times\,}
\newcommand{\phirel}{\varphi_{\rm rel}}
\newcommand{\tdrift}{t_{\rm drift}}

\begin{document}

\title{The Neural Network First-Level Hardware Track Trigger of the \belleii Experiment }

\author{S.~B\"ahr$^{1,5}$, H.~Bae$^7$, J.~Becker$^1$, M.~Bertemes$^{6}$, M.~Campajola$^8$, T.~Ferber$^1$, \\ T.~Forsthofer$^3$, S.~Hiesl$^3$, G.~Inguglia$^6$, Y.~Iwasaki$^7$, T.~J\"ulg$^4$, \\  C.~Kiesling$^3$\footnote{corresponding author, email cmk@mpp.mpg.de}, A.C.~Knoll$^4$, T.~Koga$^7$, Y.-T.~Lai$^7$,  
A.~Lenz$^{4}$, Y.~Liu$^{7}$, \\ F.~Meggendorfer$^{2,3,4}$, H.~Nakazawa$^7$, M.~Neu$^1$, J.~Schieck$^{6,6a}$, \\ E.~Schmidt$^{3}$, J.-G.~Shiu$^9$,  S.~Skambraks$^{3,4}$, K.~Unger$^1$, J.~Yin$^{10}$ 
\\
\\
$^1$ Karlsruhe Institute of Technology, Germany \\
$^2$ Physik Department, Technische Universit\"at M\"unchen, Germany
\\
$^3$ Max-Planck-Institut f\"ur Physik, M\"unchen, Germany
\\
$^4$ Chair Robotics, Artificial Intelligence (AI) and Embedded Systems, \\
Technische Universit\"at M\"unchen, Germany
\\
$^5$ now at Xilinx,
Saggart, County Dublin, Ireland
\\
$^6$ Institute of High Energy Physics, Austrian Academy of Sciences, \\A-1020 Wien, Austria\\
$^{6a}$Atominstitut, Technische Universit\"at Wien, A-1020 Wien, Austria
\\
$^7$ High Energy Accelerator Research Organization (KEK), Japan 
\\
$^8$ INFN Sezione di Napoli, Italy
\\
$^9$ National Taiwan University, Taiwan
\\
$^{10}$ Nankai University, China
}

\graphicspath{{figures/}} 
\setcounter{section}{0}

\maketitle

\begin{abstract}
We describe the principles and performance of the first-level (``L1'') hardware track trigger of \belleii, which uses the information of \belleii's Central Drift Chamber (``CDC'') and provides three-dimensional track candidates based on neural networks. The inputs to the networks are ``2D'' track candidates in the plane transverse to the electron-positron beams, obtained via Hough transforms, and selected information from the stereo layers of the CDC. The networks then provide estimates for the origin of the track candidates in direction of the colliding beams (``$z$-vertex''), as well as their polar emission angles $\theta$. Using a suitable cut $d$ on the $z$-vertices of the ``neural'' tracks allows us to identify events coming from the collision region ($z \approx 0$), and to suppress the overwhelming background from outside. Requiring $|z| < d$ for at least one neural track in an event with two or more 2D candidates will set an L1 track trigger. The networks also enable a minimum bias trigger, requiring a single 2D track candidate validated by a neural track with a momentum larger than 0.7 GeV in addition to the $|z|$ condition. We also sketch our concepts for upgrading the neural trigger in view of rising instantaneous luminosities, accompanied by increasing backgrounds.  
       
\end{abstract}
\section{Introduction}
\label{sec:intoduction}

Searches for new physics with the \belleii detector \cite{Belle2TDR}, taking data at asymmetric-energy \epem collider SuperKEKB\cite{superkekb}, will require very large integrated luminosity to challenge the predictions of the Standard Model (SM). SuperKEKB, an upgrade of the ``$B$-factory'' KEKB\cite{kekb}, is operating since 2019 and continues to produce increasing world-record luminosities, targeting 6 $\times$ 10$^{35}$ cm$^{-2}$s$^{-1}$, at the center-of-mass energy of 10.58 GeV. This collision energy corresponds to the mass of the $\Upsilon$(4S) resonance, which decays predominantly into pairs of $B$ mesons, more precisely into $B^+ B^-$ pairs or  $B^0 \, \bar{B}^0$ pairs in roughly equal numbers. In addition to the $\Upsilon$(4S) a ``continuum'' of the lighter mesons (mostly pions, kaons and $D$-mesons) is produced in the annihilation of electrons and positrons, which includes, among other less interesting quantum electrodynamics (QED) processes, the important two-body leptonic final states $\mu^+\mu^-$ and $\tau^+\tau^-$. 

All the reactions mentioned above produce final state particles which, apart from the decay products of some long-lived particles, have their origin (``vertex'') within the small collision volume of the electron-positron beams (``interaction point'' or ``IP''). The IP region has a size of order micrometers in the two directions transverse to the beam direction (``$x, y$'', or ``$r\phi$'')
\footnote{to be more precise, the collision volume in the direction perpendicular to the accelerator plane (``$y$'' direction) is only of the order of 100 nanometers, while the size in the direction``$x$'' within the accelerator plane is some tens of micrometers.} 
and a few millimeters along the beam (``$z$'') direction. There is, however, a sizeable ``background'' caused by interactions of the beam particles with the residual gas in the beam pipe (diameter 2 cm), or with the beam pipe itself, or with elements of the magnetic beam guiding and focusing system. When these interactions occur not too far from the IP, they may emit the particles produced into the \belleii detector and create signals similar to the desired annihilation events. This background is mainly characterized by particles created close to the beam line ($|x| \approx |y| \approx$ 0), while having a vertex with a large displacement from IP ($|z| \gg 0$), typically of order several cm up to a meter.  

At the IP the electron and positron bunches cross each other with a frequency of roughly 200 MHz (typically 2000  bunches of each particle type circulate in opposite directions in the rings of SuperKEKB with a circumference of 3 km). In any of the bunch crossings an interesting event might happen. Given the volume of the event data and the bunch crossing frequency of 200 MHz, it is impossible to read out the signals from the entire \belleii detector for each bunch crossing. One rather builds up a so-called ``trigger system'' which identifies, from a reduced set of detector data, those bunch crossings containing ``physically interesting'' events. 

As for the Belle experiment\cite{BelleDetector}, \belleii's trigger system has two ``levels'': The first level or ``level 1'' (L1 for short) is hard-wired and deadtime-free. It uses special fast digital detector signals which are stored in a FIFO (``$f$irst $i$n $f$irst $o$ut'') pipeline and are subjected to selection algorithms implemented in ``field-programmable gate arrays'' (FPGAs). The pipeline can hold the L1 data for 5 $\mu$s, which defines the maximum latency allowed for the L1 algorithms. There are four major detector components of the \belleii detector which contribute to the L1 trigger (see \cite{Belle2TDR} for details). These are the electromagnetic calorimeter (ECL), the central drift chamber (CDC), the K-long-muon detector (KLM) and the time-of-propagation detector (TOP). The main L1 trigger algorithms are executed with the ECL and CDC data, assisted by the KLM and TOP systems. A positive L1 decision is taken by an OR of the main trigger processors. Once an L1 trigger is asserted, the complete detector data of that corresponding bunch crossing are read out. After kinematic reconstruction of the charged and ``visible'' neutral particles in the final state, a high-level software trigger (HLT) makes the final decision and the data of the accepted events is stored on permanent media for subsequent physics analyses. One of the criteria applied by the HLT is to accept events only when the majority of the charged particles come from the IP region, i.e. $|z| < \cal{O}$(1 cm).   

\begin{figure}[t]
\begin{center}
\includegraphics[width=0.99\textwidth]{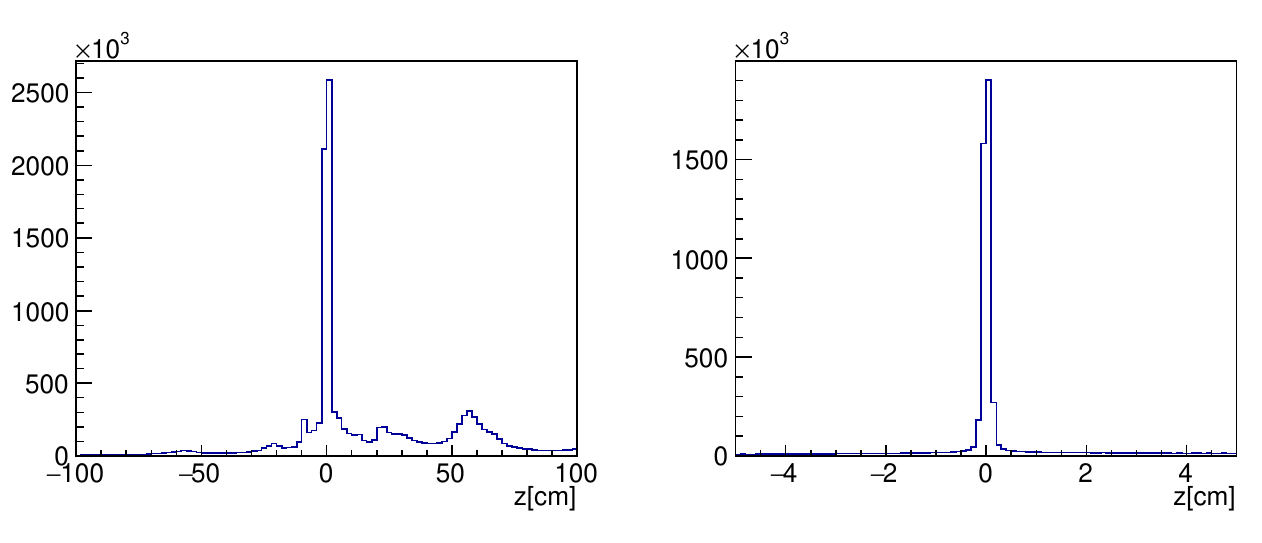}
\caption{Left: Distribution of the $z$-positions in centimeters of reconstructed tracks from \belleii, over the entire $z$ range. Right: same as left, but with higher resolution in $z$ around the IP. The events from \epem collisions contribute only about 20\% of all events. The data is from the running in the year 2020, before the launch of the \ztrig.}
\label{fig:Fig1} 
\end{center}
\end{figure}

In \belleii, the L1 trigger for charged particles (``tracks'') is derived from the CDC. In the first two years of data taking the L1 track trigger was requiring two or more tracks in the $r\phi$ plane, perpendicular to the $z$-direction of the colliding electron-positron beams. However, this ``2D'' track trigger cannot distinguish between true annihilation events ($|z|$ small) and background tracks originating far from the IP ($|z|$ large). 
Making the L1 track trigger sensitive to charged particles which originate close to the IP, while keeping the trigger rate within acceptable bounds, is of crucial importance for the efficient data taking, especially at rising instantaneous luminosities: 
An unfortunate side effect of the high luminosity is a much higher level of background, dominated by Touschek scattering~\cite{Touschek,Piwinski} and beam-gas interactions. This background produces a high rate of undesirable events with tracks mostly originating outside of IP. 
The ``problem'' of the 2D track trigger can be seen in fig.\ref{fig:Fig1}, where the majority of the events triggered come from background outside of the IP. 
On the left one clearly recognizes the ``obstacles'' for the beam approaching the IP: The peaks at about $\pm$ 60 cm are caused by the tips of the superconducting focusing quadrupoles, the excess tracks around $\pm$ 20 cm are from the two separate beam pipes for the electrons and positrons, joining into a single pipe at that distance from IP, and the peaks around $\pm$ 10 cm are from the stainless-steel cooling supports of the pixel vertex detector. 
The events in fig.\ref{fig:Fig1} have been triggered using the 2D track candidates prior to the introduction of the neural trigger. 
When the instantaneous luminosity provided by SuperKEKB surpassed the record reached by KEKB, producing also a largely increased rate from off-IP background tracks, the track trigger could no longer be maintained without an additional constraint, namely adding the third track dimension and providing an estimate for the $z$-vertex of the tracks. 

We report here on the global L1 track trigger for \belleii, using neural networks, with inputs derived from the 2D tracks to estimate their $z$-impacts along the beam direction (``\ztrig''). Due to their inherent parallel architecture, neural networks are ideally suited for solving complex pattern recognition and regression tasks within a predictable computation time, typically within fractions of a microsecond in present-day FPGAs. Furthermore, the adaptive approach of the neural trigger, being trained with real data, will ensure optimal performance under rising background conditions. The situation of backgrounds rising in parallel with SuperKEKB's program to reach its design luminosity of 6 $\times 10^{35}$ /cm$^2$s is expected to prevail for the coming years.   

This is not the first time that neural networks have been used in high energy physics experiments for event filtering at the trigger level: A level 2 neural network trigger\cite{L2NN}, performing event classification according to physics criteria, was launched in the H1 experiment at the HERA accelerator (DESY, Hamburg), and very successfully contributed to the physics of electron-proton reactions. Other attempts to use neural networks at the trigger level have been reported since, for example~\cite{cplear, dirac}, implemented in dedicated electronics. Recently, several concepts using neural networks for special triggers using FPGA hardware, still at the prototype level, have been reported~\cite{cmsl1,atlasr3} .  

The present paper is structured as follows: We will first describe the principles of the neural method for determining the origin in the 
$z$-direction of the tracks in an event. In section 2 we outline the principles of operation of the neural network trigger and in section 3 we describe the preparation of the input for the neural networks and their training. 
Section 4 presents the implementation of the neural algorithm (preprocessing and network calculations) in hardware. 
In section 5 we present the performance of the new trigger concept, in particular the minimum bias Single Track Trigger (``STT''), which requires one neural track with momentum above a certain threshold. 
This trigger was implemented in the Global Decision Logic (``GDL'') of \belleii starting in March 2021 and is now the main low-multiplicity track trigger at L1. 
In section 6 we describe ongoing developments and improved strategies for the neural trigger towards target instantaneous luminosities. A summary and outlook is given in section 7.

\section{Principle of the Neural \textbf{\textit{z}}-Trigger} 
\label{sec:principle}

As mentioned in the previous section, charged particles are triggered with the help of the CDC. The chamber, with an inner (outer) radius of $r$ = 16 (113) $\mathrm{cm}$, is equipped with 56 cylindrical layers of axial and stereo wires. The CDC contains about 15,000 sense wires, centered in drift cells with a size of 2\,cm. Six adjacent wire layers are combined to form nine so-called super-layers (``SL''). The innermost SL has eight layers with smaller (half-size) drift cells to cope with the increasing background towards smaller radii.  

The wire directions for each of the nine SLs are alternating between axial (``A'') orientation, aligned
with the solenoidal magnetic field ($z$-axis), and stereo (``U'',``V'') orientations. The stereo wires are skewed by angles between 45.4 and
74\,mrad in positive (U) and negative (V) directions with respect to the $z$-axis, thus enabling a measurement of the polar angles of the tracks.  

\begin{figure}[t]
\begin{center}
\includegraphics[width=0.7\textwidth]{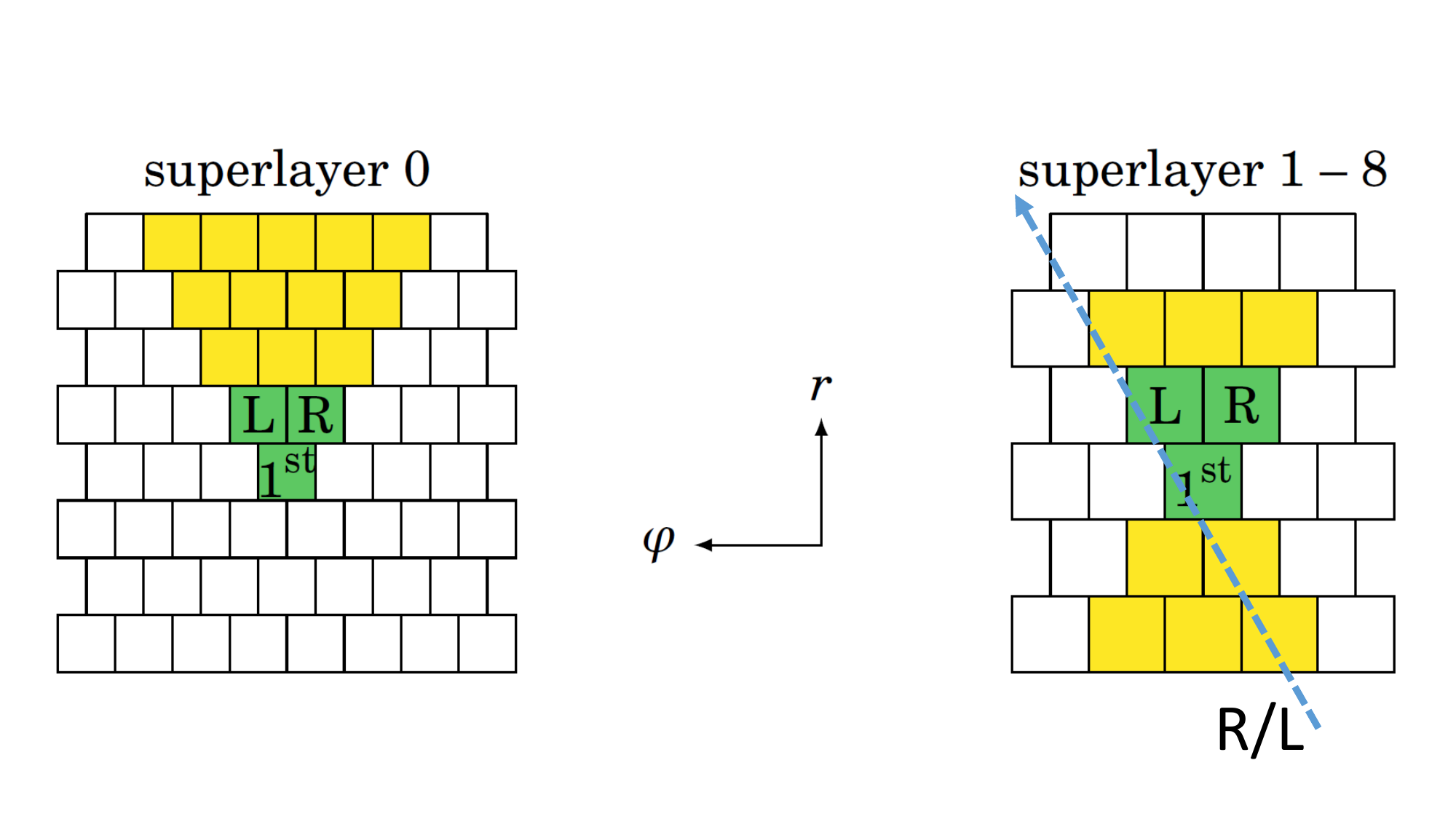}
\caption{
  Left: TS of the innermost (axial) SL of the CDC consisting of 15 drift cells, arranged in triangular shape. Right: TS in the other eight SLs consisting of 11 drift cells in hour-glass shape. The drift cells used for the TS are marked in yellow. For each TS the wire marked as ``1st'' is called the "priority wire". The two wires labeled ``R'' and ``L'' are used in case the priority wire did not fire (see text).}
\label{fig:tracksegs} 
\end{center}
\end{figure}

In contrast to offline track reconstruction using all 56 wire planes, the track finding for the trigger is based on a reduced set of sense wires in each of the nine SLs (5 axial and 4 stereo SLs). This subset of wires is selected using hard-wired so-called track segments (``TS''), which combine wires in five adjacent layers within a SL to form patterns similar to hour-glass shapes (see fig.~\ref{fig:tracksegs}, right side, for the TS in SL0 the wire set is shown on the left side). For a TS to ``fire'', its wires have to satisfy plausible patterns originating from traversing tracks. The central wire (``priority wire'') defines the spatial position of the TS. In case the central priority wire is not hit, two so-called ``2$^{nd}$ priority wires'' are defined, which take the role of the priority wire. 
 Apart from the geometrical position of the priority wires, their \drift with a resolution of 2 ns is provided by the CDC front end electronics. 
 Based on the pattern of the wires with non-zero signals within a given TS, the passing of the particle on the right or left side of the priority wire is estimated, leading to a ``signed'' \drift.

\begin{figure}[ht]
\begin{center}
\includegraphics[width=0.7\textwidth]{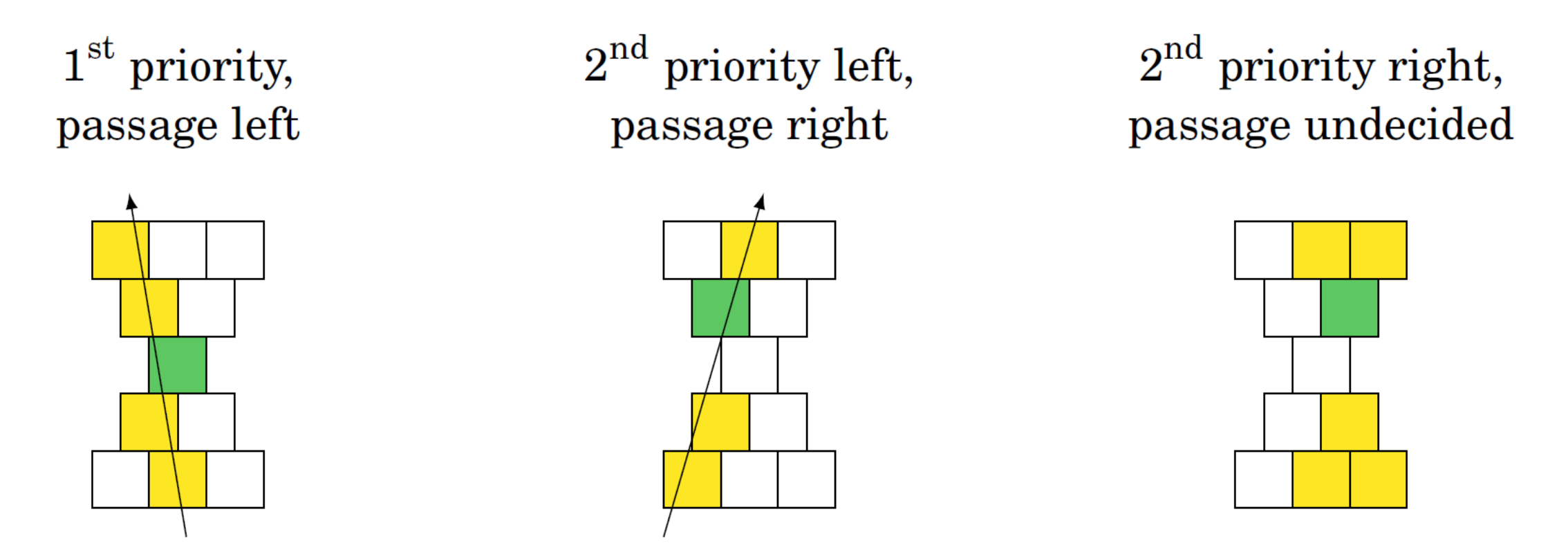}
\caption{
  Examples of hit patterns (in yellow) with different priority cells and left/right
state. The priority cell in each of the three examples is shown in green. In the case of undecided left/right passage, such as on the rightmost pattern, the \drift is set to 0, irrespective of the transmitted \drift.}
\label{fig:tsdrift} 
\end{center}
\end{figure}

Examples of wire patterns yielding the sign of the \drift are shown in fig.~\ref{fig:tsdrift}. 
Note that the patterns are determined by simple straight lines through the TS with two conditions: The angle of the line relative to the radial direction must be within the limits of $\pm 45$ degrees, and the line has to touch one of the three possible priority cells.   

Introducing the TSs also provides an extremely powerful suppression of random noise in the CDC. The set of possible wire patterns within the TSs has been encoded in look-up tables (``LUT''), which are synthesized statically during design-time. The set of priority wires in the nine SLs, including the signed \drifts, is used to reconstruct the tracks for the L1 trigger.

In the original version of the CDC track trigger the priority wires in the TSs from the five axial SLs are combined to find tracks in the plane transverse to the beam direction (``$r\phi$ plane''). These tracks are called ``2D tracks'' in the following. Using the positions of the priority wires of the axial TSs, improved by the \drift, the track finding is done using Hough transform techniques~\cite{Hough, SPohlDiss}. The Hough method assumes the track origins at ($x=0,y=0$) in the $r\phi$ plane and returns a set of 2D track candidates, which are defined by their azimuthal emission angles $\phi$ at the origin and their track curvatures $\omega = 1/R$, where $R$ is the radius of the track orbit in the $r\phi$ plane. At least 4 of the 5 axial TSs are required in the Hough transform to establish a 2D track candidate.

A seemingly natural method for a trigger processor to yield tracks in three dimensions would be to use the 2D tracks found via Hough transformation and apply track fitting algorithms by adding the priority wire information from the stereo TSs in the various SLs. However, it turned out that the combinatorics resulting from background TSs could not guarantee a fixed execution time, even using linearized track fits. Furthermore, the precision of the $z$ estimates obtained was not precise enough to provide sufficient rejection of background while retaining adequate efficiency.     

We have chosen an alternative approach for the track reconstruction in three dimensions at L1, based on artificial neural networks of the multi-layer perceptron type. Such networks are excellent candidates for the implementation of triggers in hardware: The motivation for the neural \ztrig is to provide a fast, fixed latency measurement of the $z$-vertex with sufficient precision in order to separate events from IP from those further away. A typical resolution of $\cal O$(5 cm) or better would be required to reject the dominating background from outside of the IP (see fig.~\ref{fig:Fig1}). 

In our initial studies, based on Monte-Carlo simulations~\cite{SPohlDiss, SSkamDipl, FAbudinenMaster}, the neural approach, performing a regression task, provided a sufficiently accurate estimate of the track $z$-vertex. The first step of the algorithm is to prepare the input data for the networks, choosing a set of significant variables and subjecting them to suitable preprocessing.   
Starting from the tracks found by the 2D Hough transforms, each candidate provides, in addition to the track curvature $\omega$ and the azimuthal emission angle $\phi$, the positions and signed \drifts for the priority wires in each associated TS. In a second step, a set of possible stereo TSs associated with each of the 2D tracks is selected and the stereo TS with the shortest \drift is chosen for each of the four stereo SLs.  

At present (for planned extensions see later in this paper), the inputs to the neural \ztrig are the axial TSs from the 2D track candidates and the associated stereo TSs. For each of the up to 9 priority wires three variables are calculated (``preprocessing'', see fig.~\ref{fig:TSInput} and their definition in the next section) which are fed into a single hidden layer feed-forward neural network. The two outputs of the network are the $z$ position and polar angle $\theta$ of a ``neural'' track. For a given event a number $n$ of 2D candidates will exist, leading to a number $\le n$ of neural tracks. 

The global L1 track trigger at \belleii is operated as follows: Whenever two or more 2D track candidates have been found in an event, at least one neural track with typically $|z| < 15$ cm is required for a valid L1 decision. In addition, using the estimate of the polar angle $\theta$, the momentum of each neural track is calculated. This is used to enable a trigger requiring only one 2D track candidate together with a neural track . This "Single Track Trigger (STT)" requires $|z| < 15$ cm, and a momentum $p > 0.7$ GeV. 

Since only a single neural track is required in an event, the STT is a minimum bias track trigger. The STT thus opens up the full phase space for the second and further tracks in the event, which either have very low transverse momentum ($< 250$ MeV), or are emitted with very shallow polar angles. Such tracks, although not contributing to the L1 track trigger, can nevertheless be reconstructed offline with the help of the silicon vertex detectors of \belleii. Most profit from the enlarged phase space accessible via STT triggers is expected for low multiplicity final states such as $\tau$ pair production, certain dark matter signatures, or processes like \epem $\rightarrow \pi^{+} \pi^{-} n \gamma $, ($n \ge 0)$. The cross sections for the latter processes are important to better understand the muon $g-2$ anomaly\cite{muong}.  


\section{Neural Architecture, Preprocessing, Training} 
\label{sec:training}
The neural algorithms, yielding for each track in an event an estimate for the $z$-coordinate and the polar emission angle $\phi$, are executed on Field Programmable Gate Arrays (``FPGA'') in the CDC track trigger electronic boards (see next section). Due to latency limitations at L1, only about 350 ns are available for the execution of the neural algorithm, including input variable preprocessing and delivery of the output to the GDL. Therefore, only single-hidden-layer feed-forward network architectures were considered. 

\begin{figure}[t]
\begin{center}
\includegraphics[width=0.4\textwidth]{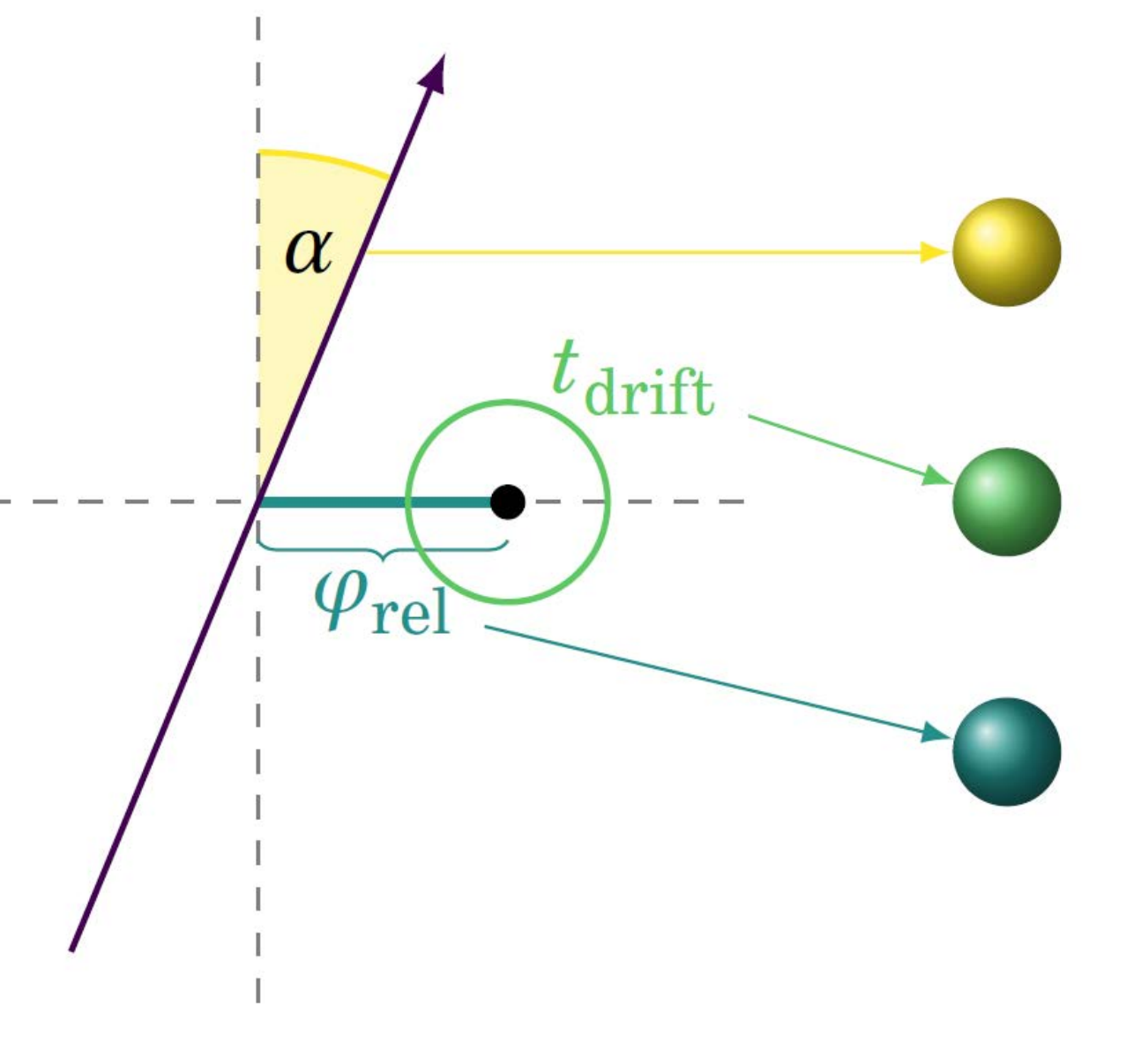}
\caption{
  Preprocessing of input variables to the neural network: for each TS contributing to a neural track, a triplet of variables is prepared, which give the crossing angle $\alpha$ of the track candidate through the TS, the angle $\phirel$ of the priority wire, and the signed \drift (see text).}
\label{fig:TSInput} 
\end{center}
\end{figure}

The inputs to the neural networks consist of triplets of variables derived for each priority wire in the associated TS for each of the nine SLs (shown in fig.~\ref{fig:TSInput}). 
The triplet of variables consists of the track crossing angle $\alpha$, the signed \drift $\tdrift$ and the relative azimuth angle $\phirel$. 
The crossing angle $\alpha$ is the inclination (or zenith) angle of the track passing the TS (see also fig.~\ref{fig:tracksegs} for illustration).
The angle $\phirel$ is given by the difference of the priority wire position in azimuth, $\phi_{\rm pw}$, and the value $\phi_{\rm extr}$ extrapolated from the Hough parameters $\phi$ and $\omega$ of the 2D track, measured at the end plate of the CDC. For the axial TSs this angle is usually close to zero. 
Then in each of the stereo SLs the TS candidates are selected using look-up tables (LUT), pre-determined from fully reconstructed tracks prior to the network training. This set of possible stereo TS candidates is defined within a range $\Delta \phi$ around the mean value $\phi_{\rm mean}$ of the two axial TS sandwiching the stereo SL. From this set the stereo TS with the shortest \drift is chosen, yielding the value $\phirel = \phi_{\rm pw} - \phi_{\rm mean}$ for the priority wire in the TS, with $\phi_{\rm pw}$ again measured at the end plate of the CDC.         

The \drifts within the individual CDC drift cells containing the priority wires are determined in the following way: The CDC front end supplies a free-running counter encoding the arrival times of the charge avalanches at the wires with a precision of 2 ns. 
The smallest value of the counters from all the CDC wires would define the time of passage $t_0$ of particles through the CDC. This $t_0$ is used, for example, in the offline reconstruction. 
 
 Due to the limited number of wires for the trigger, however, no global $t_0$ is available. We have therefore chosen the method of ``self timing'', looking for the smallest counter value $t_{\rm min}$ in the set of priority wires associated with the neural track candidate. 
 This value is then subtracted from the timing counters in all the associated priority wires, yielding the unsigned \drifts. The right-left (R/L) ambiguities for the \drifts are lifted again by pre-determined LUT, analyzing the hit patterns in the TSs from a set of simulated tracks spanning a wide range  of momenta and emission angles. 
 Note that the patterns do not depend on the $z$-origin of the tracks. Whenever a clear decision can be made for the track passing right (left) from the priority wire, the drift time is assigned to a  negative (positive) value (see fig.~\ref{fig:tracksegs}). 
 In the case no decision can be made, the \drift is set to zero (``track passing close to the wire''), irrespective of the size of the \drift. 
The third input of the triplet is the crossing angle $\alpha$ of the track through the TS. This quantity is derived from the Hough maxima $\phi$ and $\omega$ and the known radial position of the priority wires. 

\begin{figure}[t]
\begin{center}
\includegraphics[width=0.9\textwidth]{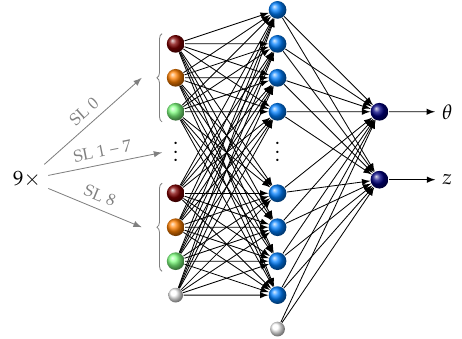}
\caption{
  Sketch of chosen architecture for the fully connected feed-forward neural network. The input consists of 9 triplets of data from the 9 SLs, giving a total of 27 inputs. Due to limitations in the latency only 81 nodes in a single hidden layer are implemented. Two output nodes provide the estimates for the $z$ origin of the track and its polar emission angle $\theta$.}
\label{fig:NetArchitecture} 
\end{center}
\end{figure}

Since there are nine SLs (5 axial and 4 stereo) the number of inputs is $3 \cdot 9 = 27$. The number of neurons in the hidden layer and the bit widths of the synaptic connections are determined by the available resources of the FPGAs. 
An optimum of 12 bits was found with 81 hidden nodes for the synaptic connections (more details see the hardware section). 
A schematic of the fully connected multi-layer-perceptron architecture is shown in fig.~\ref{fig:NetArchitecture}. Each line from the colored nodes represents a weight, while the lines from white nodes indicate the offsets, as determined by the training process. Details on the encoding into fixed-point numbers are given in the next section. 

The training of the networks is done with the target values $z_{\rm rec}$ and $\theta_{\rm rec}$ taken from the fully reconstructed tracks from an event sample, fulfilling any of the L1 trigger conditions from the CDC or ECL recorded during the data taking. 
Since not always all TSs in the four stereo SLs are carrying signals (e.g. due to local inefficiencies in the CDC, or tracks missing the innermost or outermost stereo SLs), four additional networks, each one for a specific missing stereo SL, are trained. 
Note that at least 3 out of the 4 stereo SLs are required to create a valid neuro track. 
It was found that these four networks (we call them ``experts''), each associated with one of the missing stereo SLs trained separately, are performing better than a single network with all cases included. 
A total of 5 networks, depending on the set of stereo TS in the data samples, are trained (``expert 0'' to ``expert 4''). 
For each of the 5 sets the same loss function $\mathcal{L}$ is defined: 
$$
\mathcal{L} = \sum{((z_{{\rm rec},j} - z_{{\rm net},j})^2 + (\theta_{{\rm rec},j} - \theta_{{\rm net},j})^2)}/N
$$
where the sum runs over the $N$ tracks in the training sample. For numerical stability, all input variables as well as the outputs are rescaled to the norm interval $[-1, +1]$ and the activation function is chosen as the hyperbolic tangent.

For the training of the first deployment of the neural trigger in the fall of 2020, the data collected during the spring of the year 2020 has been taken. At that time, the instantaneous luminosity was around $1.5 \times 10^{34}$ cm$^{-2}$  s$^{-1}$. The training software used then was from the FANN library included in the basf2 software framework~\cite{basf2}. Typical $z$-resolutions around 5 cm were obtained from the initial training with the FANN library. This turned out to be sufficient for rejecting a large fraction of the off-IP events during the data taking in the year 2021 (see Sect.~\ref{sec:performance} below).

With rising luminosity and rising backgrounds towards the end of 2021 a new training was launched, based on the PyTorch library~\cite{pytorch}. 
Here, we also include a second norm weight regularization into the loss function to punish excessively large weights and thus avoid overfitting issues.
The loss function is minimized by Adam \cite{adam}, a first order optimizer, using mini-batches. 
The gradients are calculated with the backpropagation algorithm implemented in PyTorch. Convergence is controlled by an independent validation data set. 
The sample sizes for training and convergence tests are generally of order 300 k tracks. 
The training itself is done in floating point arithmetic. 
The transformation of all the inputs, outputs and network parameters to integers for the computations on the FPGAs is described in Sect.~\ref{sec:hardware}. 
The result was a $z$-resolution of about a factor two better compared to the FANN training. Details on the performance of the new networks are given in Sect.~\ref{sec:performance}.

We should mention here that the preprocessing of input variables as well as the operation of the five networks with their parameter sets are implemented in the basf2 software framework. 
The  simulation uses the full trigger information (48 time slices of the pipeline) recorded for each 256th event. 
With this information we can simulate and monitor the performance of the hardware in a subset of the data and also estimate the expected performance of alternative network architectures.


\section{Hardware Realization} 
\label{sec:hardware}

The neural networks form the last part of the CDC track trigger system, which imposes the requirement of being fully pipelined. 
The implementation of the neural algorithms and the preprocessing of the input variables in FPGA  hardware~\cite{SBaehrDiss} for the \ztrig are outlined in the following. 

\subsection{Integration and Requirements}

\begin{figure}[t]
	\centering
	\includegraphics[width=\linewidth]{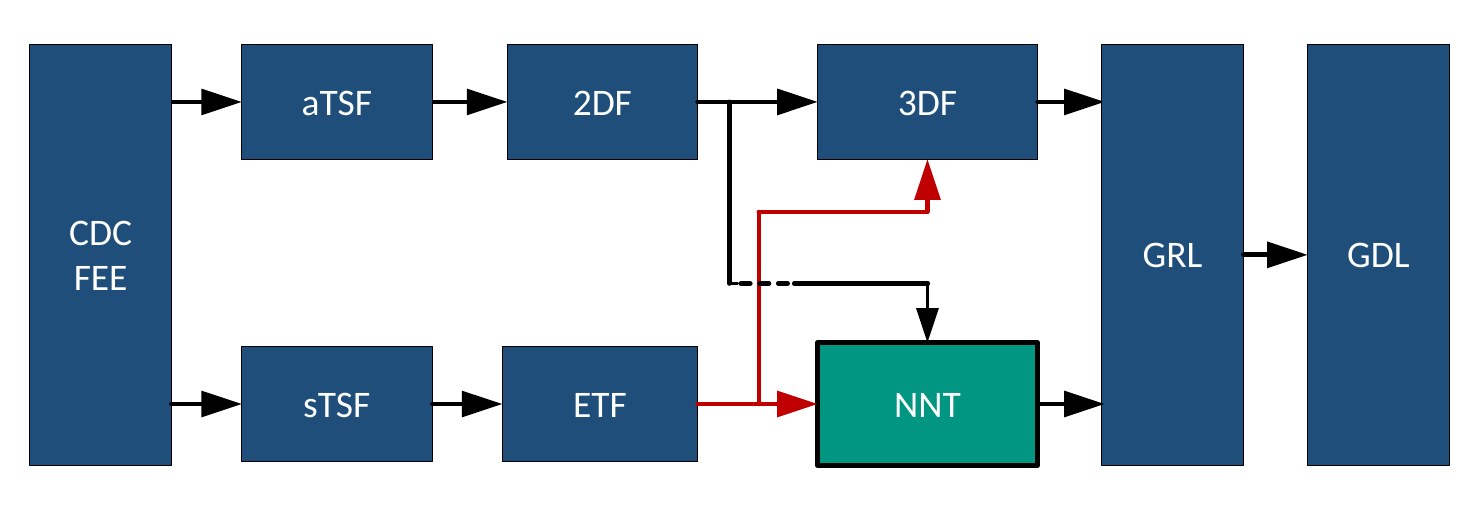}\caption{Integration of the neural trigger into the CDC trigger system. The various units are explained in the text. The unit ``3DF'', short for ``3D fitter'', has not been implemented and is therefore not discussed further.}
	\label{fig:hw:cdcarch}
\end{figure}

The location of the neural trigger in the CDC trigger system is shown in fig.~\ref{fig:hw:cdcarch}. 
The neural part is making ample use of the preceding processing modules which are organizing the wire information from the CDC front end electronics (CDC FEE) for further processing. 
To minimize the overall latency, the wire data are split in geometrical quadrants in the $(x,y)$ plane transverse to the beams, and the processing is done in parallel for all four quadrants. 
At the first stage, the track segment finders (aTSF, sTSF) concentrate the wire information in the axial TSs and stereo TSs (see previous sections). The axial TSs are used by the 2D finder (2DF) and build the 2D track parameters using Hough transforms, as described in the previous section. 
These parameters are sent together with their associated TSs to the neural trigger system (NNT). 
The stereo TSs on the other hand are sent directly to the neural trigger. 
There they have to be combined and related to the 2D candidate tracks. 
The selection of the related stereo segments is performed by an internal hit selection module. 
One of the inputs generated for each TS and directly used by the network is the \drift. 
In the original concept of the trigger, its calculation is based on an event time provided 
by the event time finder sub-system (ETF). 
However, this sub-system was not available at the time. To compensate for this, an internal event time is calculated by setting the earliest priority time, out of all related TSs, as the event $t_0$ (see preceding section).\\

The allowed latency of the neural trigger is determined by the latency of the 2D track finder, as it provides its input later than other systems, and the arrival deadline set by the GRL(``Global Reconstruction Logic'')/GDL in order to make the decision in time. 
The major contribution to the delay is generated by the communication between the sub-systems involved, which adds about 300-400 ns per transmission (see the full lines in fig.~\ref{fig:hw:cdcarch}). 
Taking into account that the neural trigger has to execute the transmission twice, i.e. input from the 2D finder and output to the GRL, a latency budget of only about 300 ns is available for the entire neural processing chain (the total latency of the \belleii trigger system is about 5 $\mu$s). 
The input is transmitted at 32 MHz (``trigger clock''), which should be matched for the output to fully keep up with the incoming rate. 
However, it is rarely fully occupied, thus processing of successive tracks is deferred at the trigger in case the pipeline is currently occupied. 
This should only be problematic for the latency in cases of a high number of inputs, i.e. when more than eight successive clock cycles are arriving in successive transmission cycles.   

Due to limited resources, it is currently only possible to process and transmit to the GDL at most one track per clock cycle per quadrant. 
If two tracks in the same quadrant arrive in the same cycle, the second one, ordered by momentum, is suppressed. 
The second track can, however, be buffered and deferred. 
But this architecture detail was thus far not used in the experiment, because it did not affect the performance of the neural trigger.

The last aspect of integration is the choice of the FPGA to be used. Initially, the trigger was intended to operate on the custom-designed UT3 platform. 
Due to the restricted stock of FPGAs at that time, it had to be designed for and implemented on the smaller XC6VHX380T instead of the larger XC6VHX565T. 
Future ongoing designs for the UT4 platform are not considered in this note. 

\subsection{General Architecture and Implementation}

\begin{figure}[t]
	\centering
	\includegraphics[width=\linewidth]{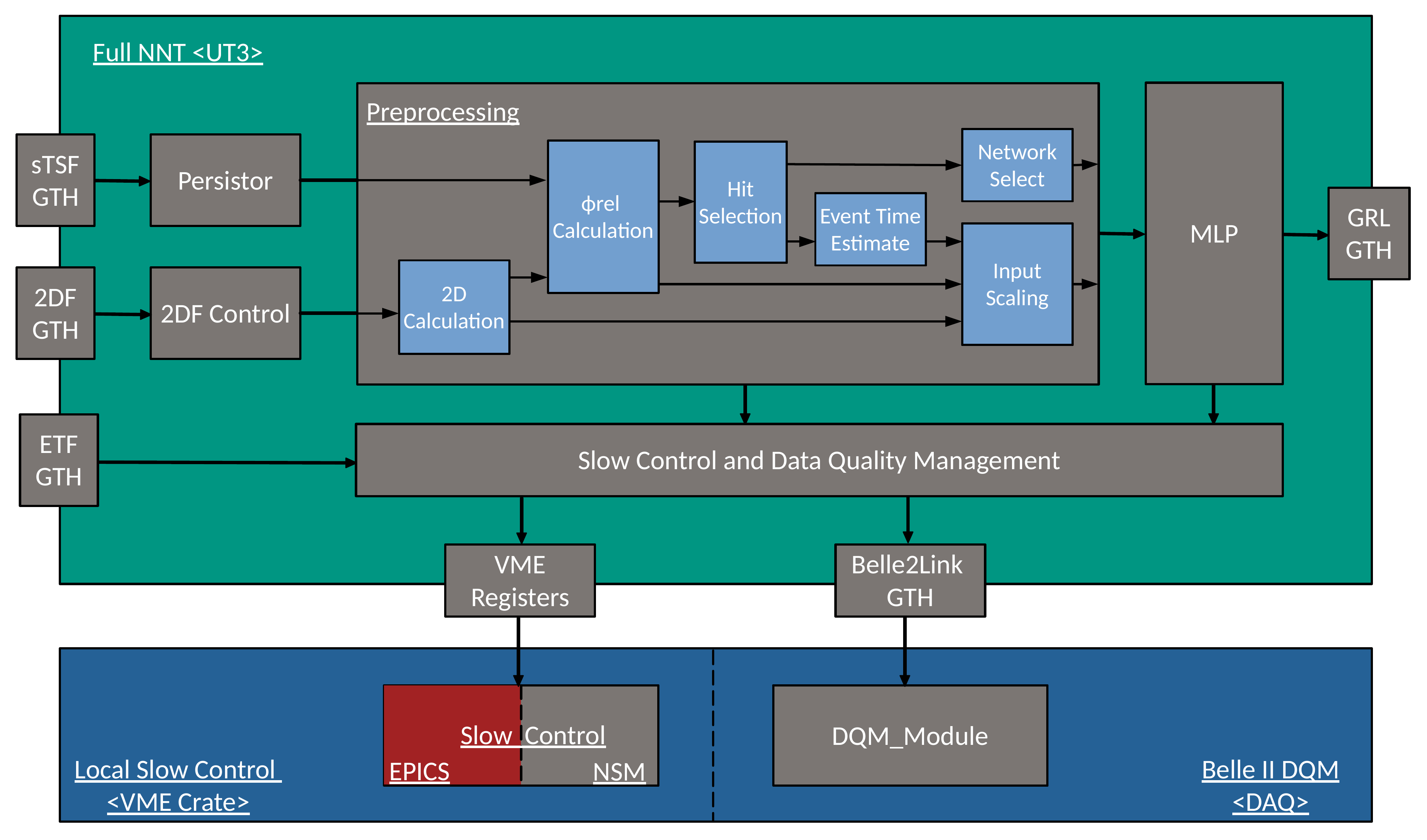}\caption{Details of the overall architecture of the neural trigger hardware.}
	\label{fig:hw:complete}
\end{figure}

Details of the neural system architecture are shown in fig.~\ref{fig:hw:complete}. 
The architecture can be partitioned into input handling, processing, and monitoring. Input handling is responsible for implementing the protocol of the input sources and merging the separate data streams into one combined stream for the subsequent processing chain. 
The key components here are the persistor and the unsyncher. 
The unsyncher is buffering stereo TSs and 2D data to achieve a constant timing offset. The persistor module is tasked with generating pools of viable TS candidates to be considered for the track estimation. 
The input handling is followed by the preprocessing and the realization of the neural network. A network memory with decision logic is placed in between the two to decide on the weight set to be loaded depending on the current input data, i.e. which expert network needs to be used. 
All processing modules are implemented to provide parameterization of the essential characteristics. These are bit widths, degree of parallelism, number of register stages, and choice of implementation between SRAM and DSPs. In addition they are designed to allow the application of retiming in order to keep signal delays between registers balanced. 
This requires compliance with the design guides provided by the tool vendors.

\textit{\textbf{Preprocessing}}: The general structure of the preprocessing as shown in fig.~\ref{fig:hw:complete} consists of several data paths which are executed in parallel to keep the latency low. The individual stages are internally pipelined to achieve the frequency goal of 127 MHz. Differences in individual processing latencies are compensated by shift registers buffering intermediate results; those are not explicitly shown in the architecture. 
With all inputs from the TSF and 2DF the preprocessing is generating control signals which will load a specific weight set into the neural network module. In the base configuration, the weight set will be loaded depending on the presence of stereo layers with suitable TSs (experts). 
The scheduling of the individual operations is performed as soon as possible under the present data dependencies, with the input scaling being the common point of synchronization right before the entry to the neural network. 
In general, across all modules the best choice for implementation was to use SRAM instead of DSPs. Due to their location relative to the transceivers, high availability, flexible routability, together with the high demand of DSPs within the neural network, it was more economic, in terms of resources and latency, to avoid DSPs in the preprocessing steps by using tool directives instead. 

\textit{\textbf{Neural Network}}: Processing in the neural network is performed by artificial neurons. These can be implemented by combining multiply-and-accumulate operations with the chosen activation function. 
The remaining parameters are then the resources to be used for the operations and the schedule for individual operations. In general, the best performance can be achieved on FPGAs when using DSPs for multiplications with variable input values. As such this realization is relying on DSP units. 
In addition, these units are already equipped with an internal accumulator which can be used to further increase efficiency. Its usage is, however, determined by the chosen schedule. First of all the amount of operations to be scheduled, namely $(27+1) \cdot 81+(81+1) \cdot2=2432$ operations (see fig.~\ref{fig:NetArchitecture}), is exceeding by far the amount of DSPs available on the FPGA in the UT3, which are limited to about $900$ units. 
Due to the complexity in routing, the input signals through the physical location of the DSP on the floor plan of the FPGA, it is furthermore not possible to have an architecture utilizing all the available resources. Experimental studies have shown that the FPGA architecture  does not support a utilization above 65 \%, considering the system clock of 127 MHz.

To optimize resource utilization, we use the technique of multiplexing. This involves deferring a subset of inputs to later clock cycles during input multiplication, thereby reducing the resource demand per clock cycle. The positive side effect of this is that the internal accumulator can now be used to produce partial sums without the need for additional adder logic. The question now is how to determine the factor of multiplexing, and the choice was to define it depending on the resulting resource consumption. 
A multiplexing factor of $5$, which means $5$ inputs are processed on one DSP unit across multiple clock cycles, yielded the desired resource consumption which in turn resulted in timing closure later on. 
We note here that there are several more advanced approaches to increase the efficiency of using DSPs, for example SIMD style operations with multiple inputs and weights being interleaved to make use of the 18 and 21 bit wide data ports of the units. 
However, the choice here was to increase the bit width to the maximum instead of increasing operational performance itself in order to achieve higher resolutions. 
There is, on the other hand, a trade-off here between free resources and resolution that could be further explored. The activation function (hyperbolic tangent) is realized with a LUT which includes logic to negate the result, and different resolutions to efficiently implement the function. Here we use the Dual-Port logic provided by BRAM units on the FPGA to reduce resource consumption by sharing a LUT for two artificial neurons.

\begin{figure}[t]
	\centering
	\includegraphics[width=\linewidth]{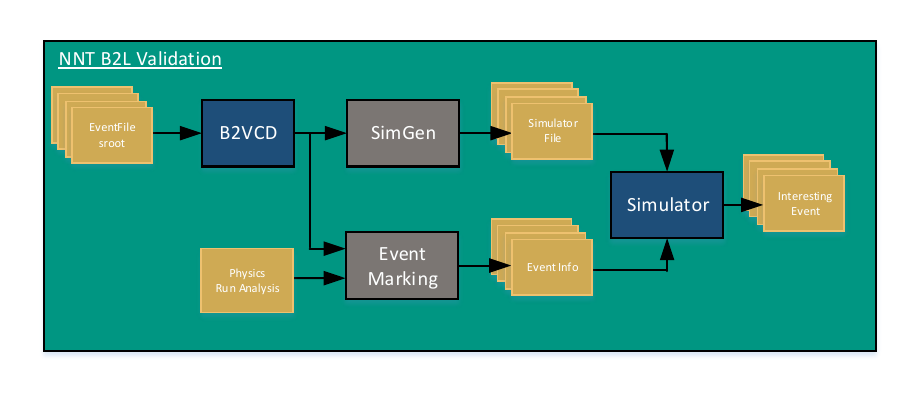}\caption{Process for offline verification of the hardware on clock cycle granularity.}	
	\label{fig:hw:verification}
\end{figure}

\subsection{Configuration and Testing}
Verification of the hardware is based on the usage of B2Link data~\cite{belle2link} in combination with the tool B2VCD. This tool is unpacking the data from specified boards and transforms integers into a VCD format which can then be read by signal plotting tools such as gtkwave~\cite{gtkwave} for subsequent analysis.
The file can additionally be used to generate stimuli for a cycle-by-cycle recreation of the encountered scenario in an HDL simulation tool. 
This is performed in the SimGen tool~\cite{simgen} which in addition incorporates the IO specifics of the NNT to map data onto the IO interfaces that are used at HDL-level. 

With this setup it is already possible to validate the runtime hardware against functional simulation to uncover problems originating from signal delays or improper interfacing. 
In addition to this an event-marking tool was implemented with the task to combine offline analysis which compares the hardware with ideal software execution and low-level HDL simulation. 
Here the idea is to identify problematic events after the run and mark down the position in the clock-by-clock simulation.

\begin{table}[ht]
	\begin{center}
		\begin{tabular}{|c|c|c|c|c|c|}
			\hline
			TS Depth & Persistence & $\alpha$ BW & $\phi_{rel}$ BW & Hit Selection & Multiplication  \\\hline
			24 & 16 & 14 & 24 & 1 stage & Slices  \\\hline
		\end{tabular}
		\caption{Architecture configuration of the preprocessing.}	
		\label{tab:hw:prepro_config}
	\end{center}
\end{table}
\begin{table}[!ht]
	\begin{center}
		\begin{tabular}{|c|c|c|c|c|c|}
			\hline
			Weight Sets & Weight BW & Input BW & MAC & Activation & MUX  \\\hline
			5 & 18 & 13 & DSP & LUT full optimized & 5 \\\hline
		\end{tabular}
		\caption{Architecture configuration of the neural network.}	
		\label{tab:hw:full_mlp_config}
	\end{center}
\end{table}
\begin{table}[!ht]
	\begin{center}
		\begin{tabular}{|c|c|c|c|c|c|}
			\hline
			Slices & Registers & DSPs & BRAM & Frequency & Latency  \\\hline
			46\% & 14\% & 53\% & 49\% & 127 MHz &   $288ns$ \\\hline
		\end{tabular}
		\caption{Implementation characteristics for the full setup trigger.}	
		\label{tab:hw:full}
	\end{center}
\end{table}

The hardware configuration and implementation characteristics are discussed in the following. The configuration of the preprocessing is shown in Table~\ref{tab:hw:prepro_config}, and for the neural network in Table~\ref{tab:hw:full_mlp_config}. Here, TS depth is referring to the maximum amount of stereo TSs which are buffered at any clock cycle, higher depths would exceed the timing and resource budget. Persistence describes the number of clock cycles for which a segment is buffered before being invalidated. The columns with ``BW'' are describing the internal bit widths for the $\alpha$ and $\varphi_{rel}$ values that are internally used, and ``MAC'' stands for Multiply-accumulate, ``MUX'' gives the multiplexing level. Table~\ref{tab:hw:full} is showing the resulting latency, frequency, and resource characteristics of the fully integrated trigger for the V6VHX380T FPGA on the UT3. They are fulfilling all the timing requirements and resource constraints.


\section{Performance of the Neural $\mathbf{z}$-Trigger} 
\label{sec:performance}

In this section we present the performance of the \ztrig, based on the data from its launch in early 2021 until the beginning of the Long Shutdown (``LS1'') in June 2022. The \ztrig was commissioned in late 2020, being monitored but not yet activated at that time. From early 2021 onward the neural trigger was activated and operated as a global track trigger at L1, with the requirement that at least one neural track is found in the event obeying the condition $|z| < d$. Typical values of $d$ are 20 cm or less. For the performance studies we will compare the neural tracks (``neuro track'') with the corresponding fully offline reconstructed tracks (``reco track'').

\begin{figure}[!htp]
\begin{center}
\begin{minipage}[]{0.48\linewidth}
\centering
\includegraphics[width=0.95\textwidth]{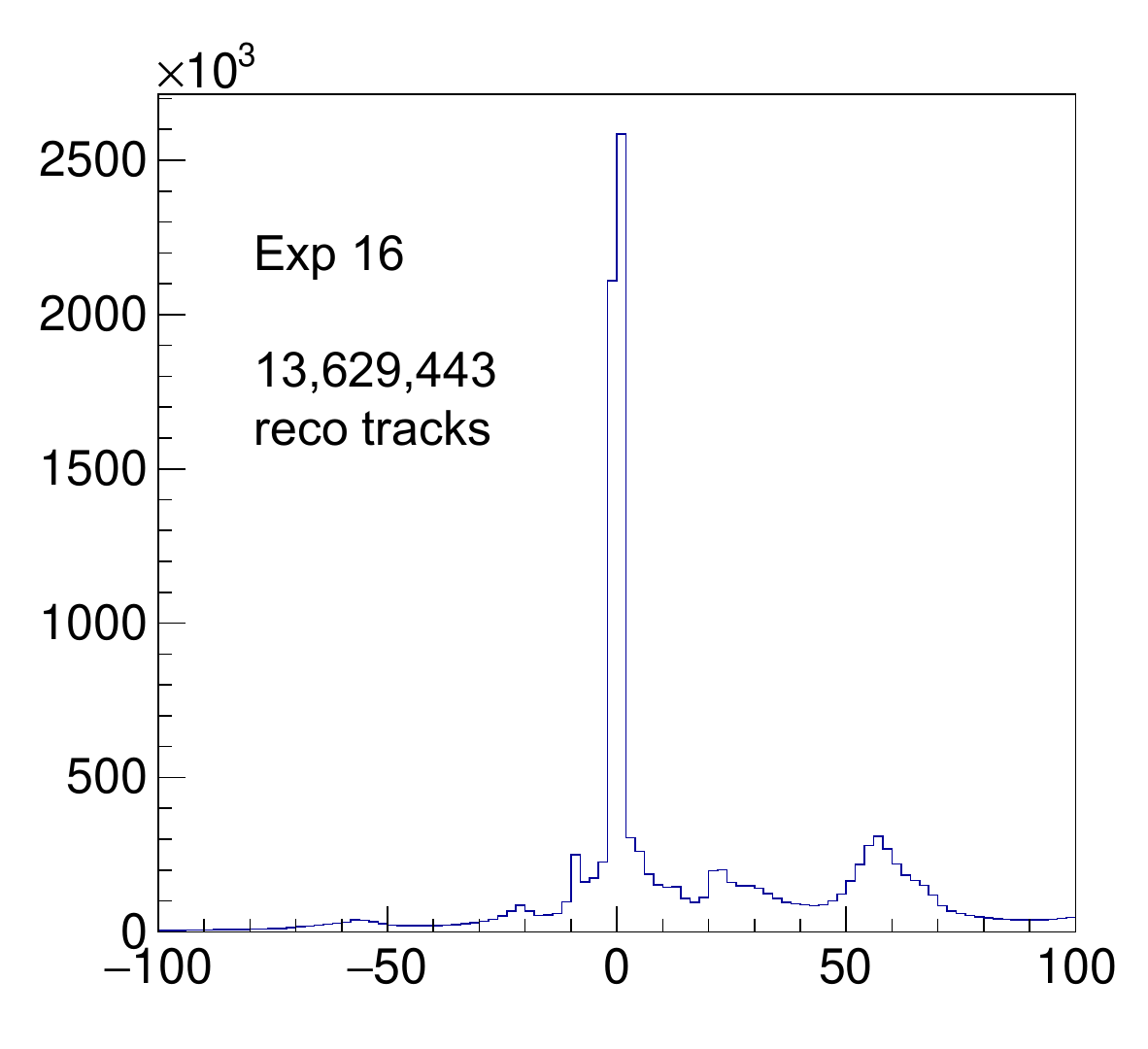}
\includegraphics[width=0.95\textwidth]{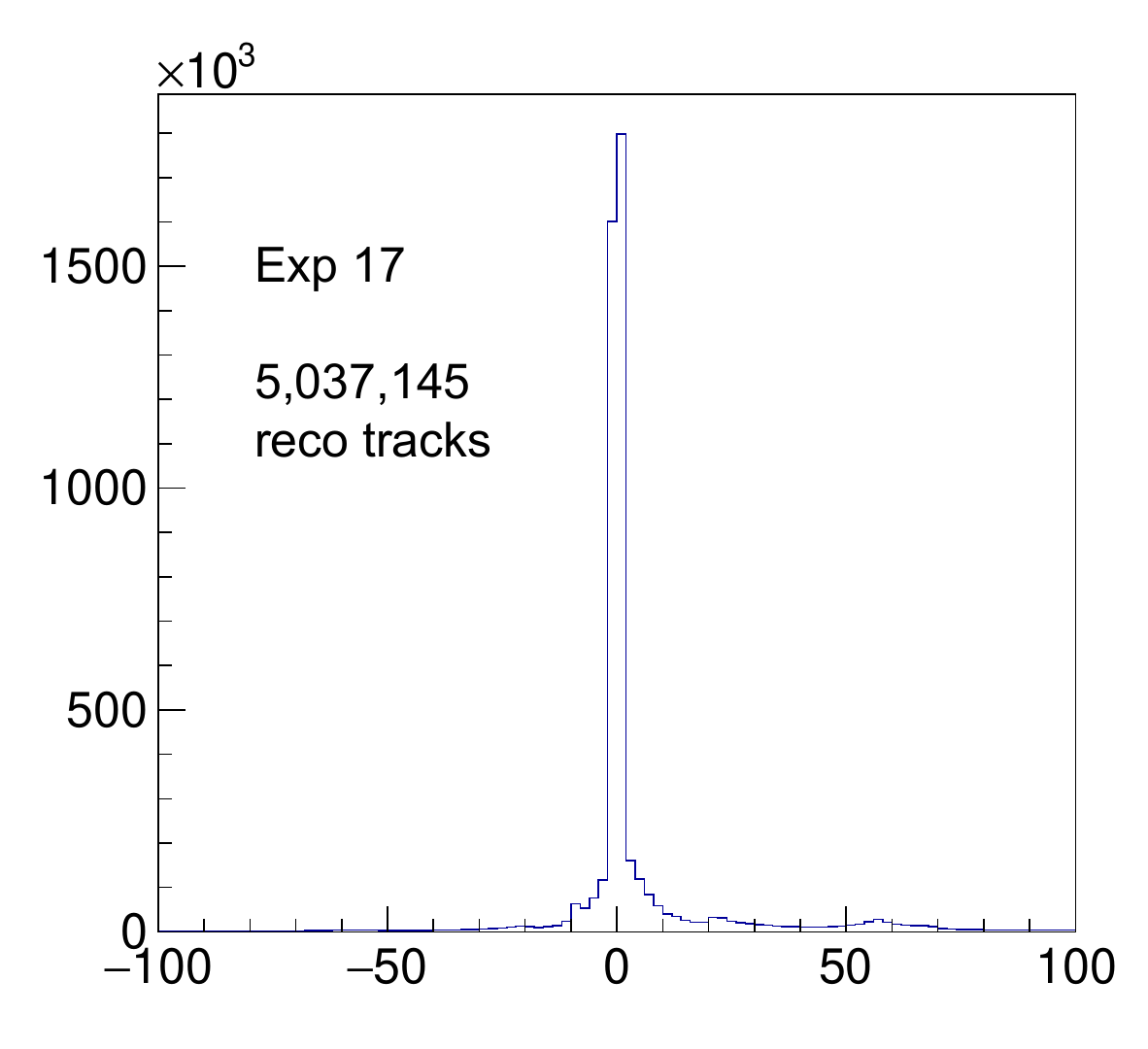}
\includegraphics[width=0.95\textwidth]{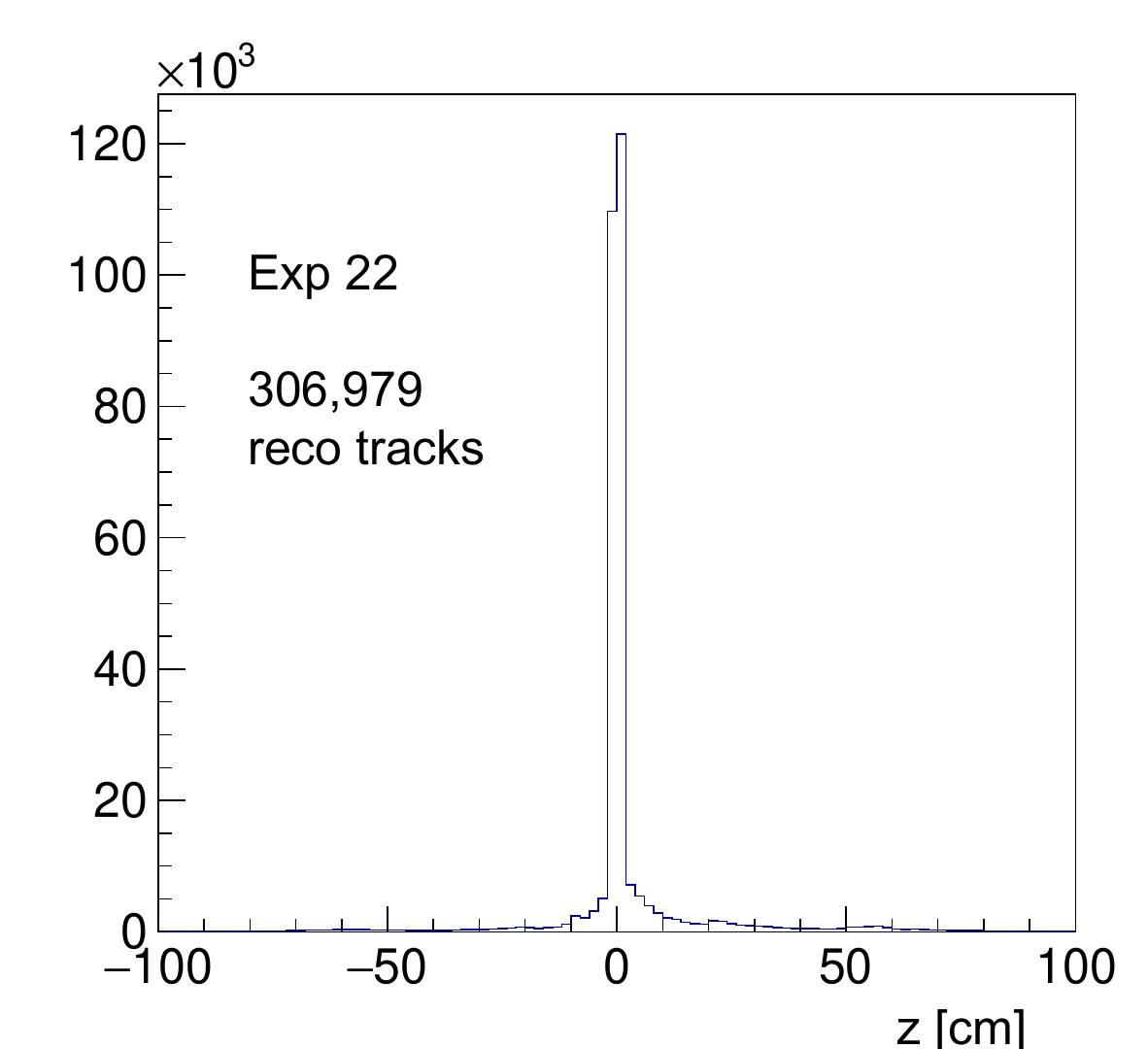}
\end{minipage}
\hfill
\begin{minipage}[]{0.48\linewidth}
\centering
\includegraphics[width=0.95\textwidth]{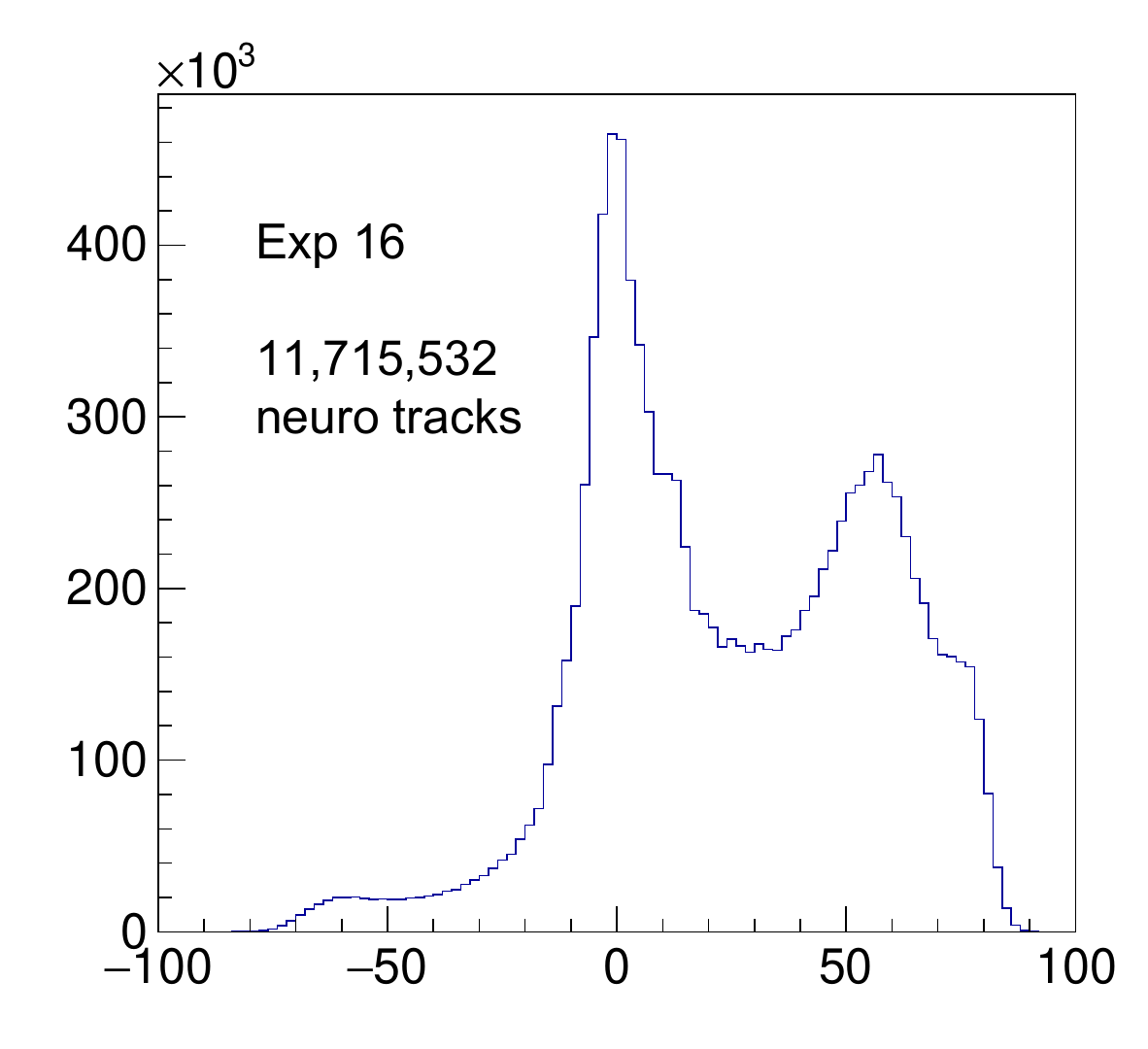}
\includegraphics[width=0.95\textwidth]{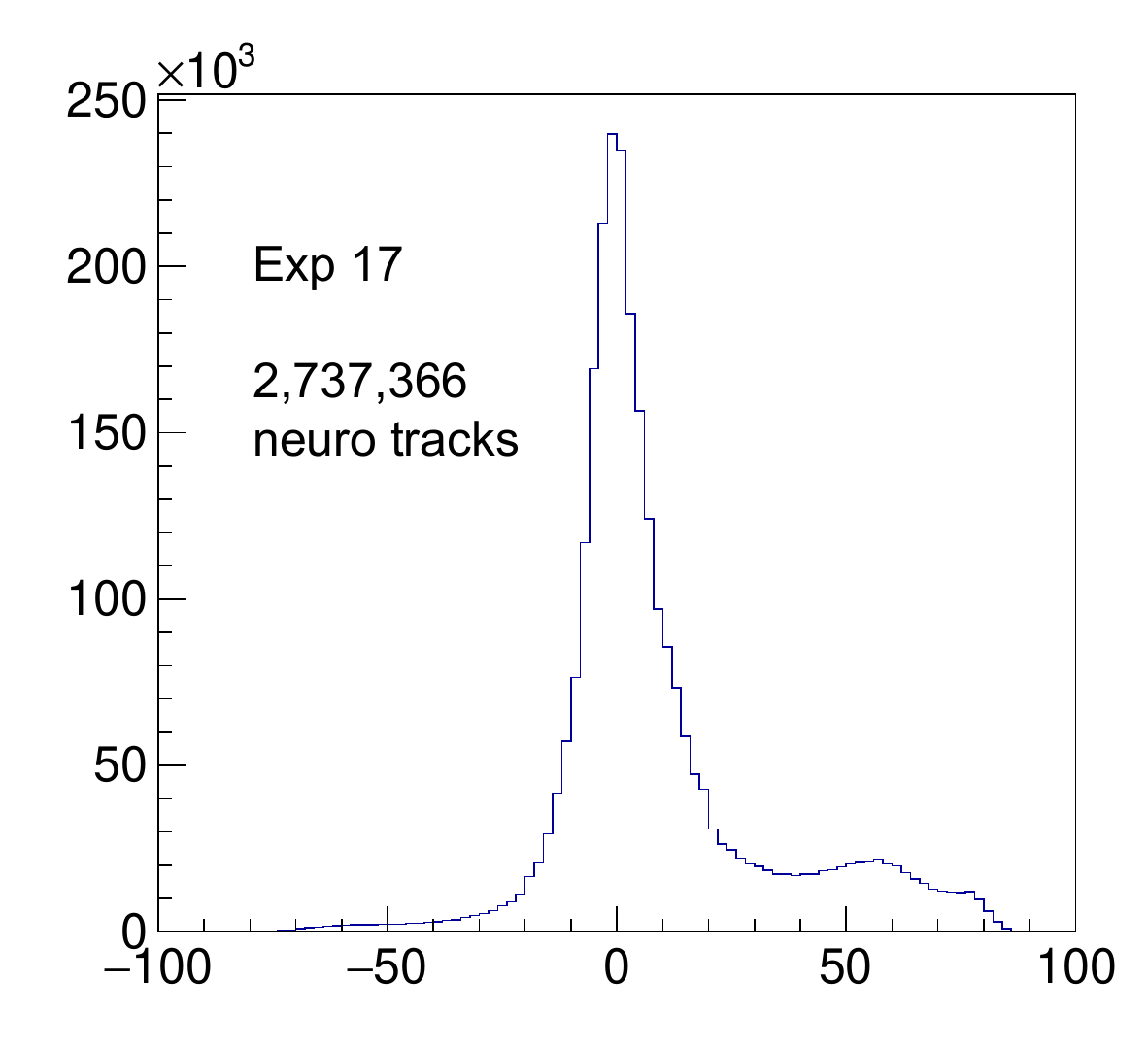}
\includegraphics[width=0.95\textwidth]{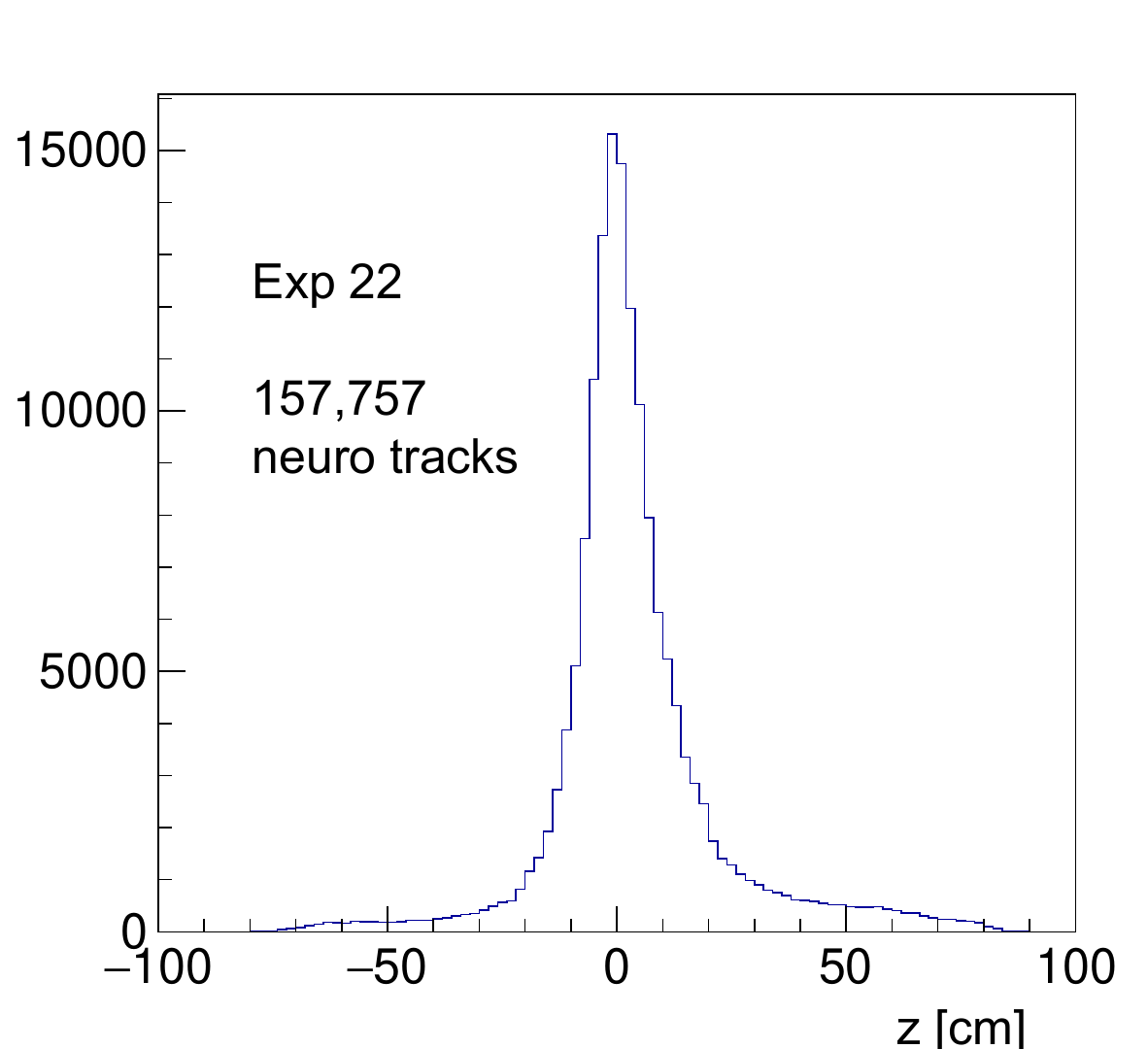}
\end{minipage}
\caption{
  Left column: Distributions of fully reconstructed tracks using the offline reconstruction software package. Right column: Distributions of found neural tracks. Upper row: Exp.~16, the neural $z$-trigger was not yet enabled. Middle row: Exp.~17 with $z$-trigger enabled for this and the following experiments. Lower row: Exp.~22 (end of 2021 running). In this latter period the machine background was significantly reduced (see text).}   
\label{fig:allzplots} 
\end{center}
\end{figure}

The most important characteristic of the neural approach is the achieved resolution in the $z$-impact of the track, as this quantity is subject to cuts in the GDL for the event acceptance at L1. 
Before going into the details of a $z$-dependent resolution for the neuro tracks, we show in fig.~\ref{fig:allzplots} the distributions of the $z$-vertices along the beam axis for reco tracks (left column) and for the tracks found by the neural trigger hardware (right column). 
The three rows correspond to different data taking periods, labelled by increasing `` experiment numbers'': Exp. 16 (top row), Exp. 17 (middle row), and Exp. 22 (bottom row).
The data come from special streams written to tape, sampling all L1 triggers with complete pipeline information over the entire experiments. 
Typical running periods for these experiments are in the order of a month or more. The different experiments are characterized, for example, by special settings of the machine parameters and by certain combinations in the trigger menus. 

All results from the \ztrig (right column of fig.~\ref{fig:allzplots}) have been obtained from networks with the initial FANN training. 
No selection of events or tracks has been made, neither in the reco tracks nor in the neuro tracks found at L1. 
As can be seen from the figures, the number of neuro tracks is smaller compared to the number of reco tracks. This is expected due to the limited polar angle acceptance of the neuro tracks, and tracks with transverse momentum less than 250 MeV necessary to pass the required minimum of four of the five axial wire planes.
Furthermore, some reco tracks are reconstructed offline solely in the vertex detectors, with a minimum of CDC wire planes, insufficient to set a wire trigger. 
Such reco tracks typically correspond to charged particles with very shallow polar emission angle or very low transverse momentum. Concerning the $z$-resolution of the neuro tracks, a first look at the peak at $z = 0$ shows a resolution definitely better than 10 cm. 
This demonstrates the high potential of the \ztrig to suppress background events from outside of IP. 

Comparing the distributions in the first row of fig.~\ref{fig:allzplots} to the ones below one observes a clear difference in the fraction of tracks coming from regions outside of the interaction point (IP): In Exp.~16 (top row, beginning of the luminosity runs of the year 2021) the \ztrig was not yet activated. The charged particles were triggered requiring two or more 2D track candidates. 
In this experiment, the machine background was already substantially larger compared to the running in 2020. 

Since the total trigger rate, dominated by the track trigger, was close to the DAQ limit at that time (early 2021), the decision was taken to switch the \ztrig to active (Exp.~17, middle row). 
For the \ztrig to fire, at least one neuro track with a conservative cut of $|z|<20$ cm was required. The strong rejection of events from outside of the IP is evident. However, one also notices neuro tracks with $|z|$-values larger than 20 cm in the plots. The reason for this are events with more than one neuro track, where one (or more) of these tracks fulfills the $z$ cut. This fires the \ztrig, but there may still be other neuro tracks in these events having larger displacements. These then enter the plots on the right side as well. 

When the \ztrig was switched on, the track trigger rate was reduced by roughly a factor of 2, keeping the total L1 rate well below the required DAQ limit and the track trigger could from then on be operated for physics data taking without prescale.

During the fall of 2021 the instantaneous luminosity was steadily increased, accompanied by a strong increase of background. 
With the beam currents raised above 1000 mA (800 mA) for the positrons (electrons), the luminosity had reached a new record of $3.8 \times 10^{34}$ \cmss.     
However, there were also some shorter run periods where the background conditions for the trigger were less severe, although the instantaneous luminosity did not change significantly. As an example, the data from the end of the data taking (December 2021) is shown in the last row of fig.~\ref{fig:allzplots}. Here one observes a reduced contribution from large $z$, caused by the favorable background conditions in the SuperKEKB accelerator.

\begin{figure}[!hbt]
\begin{minipage}[]{1.0\linewidth}
\centering
\includegraphics[width=0.48\textwidth]
{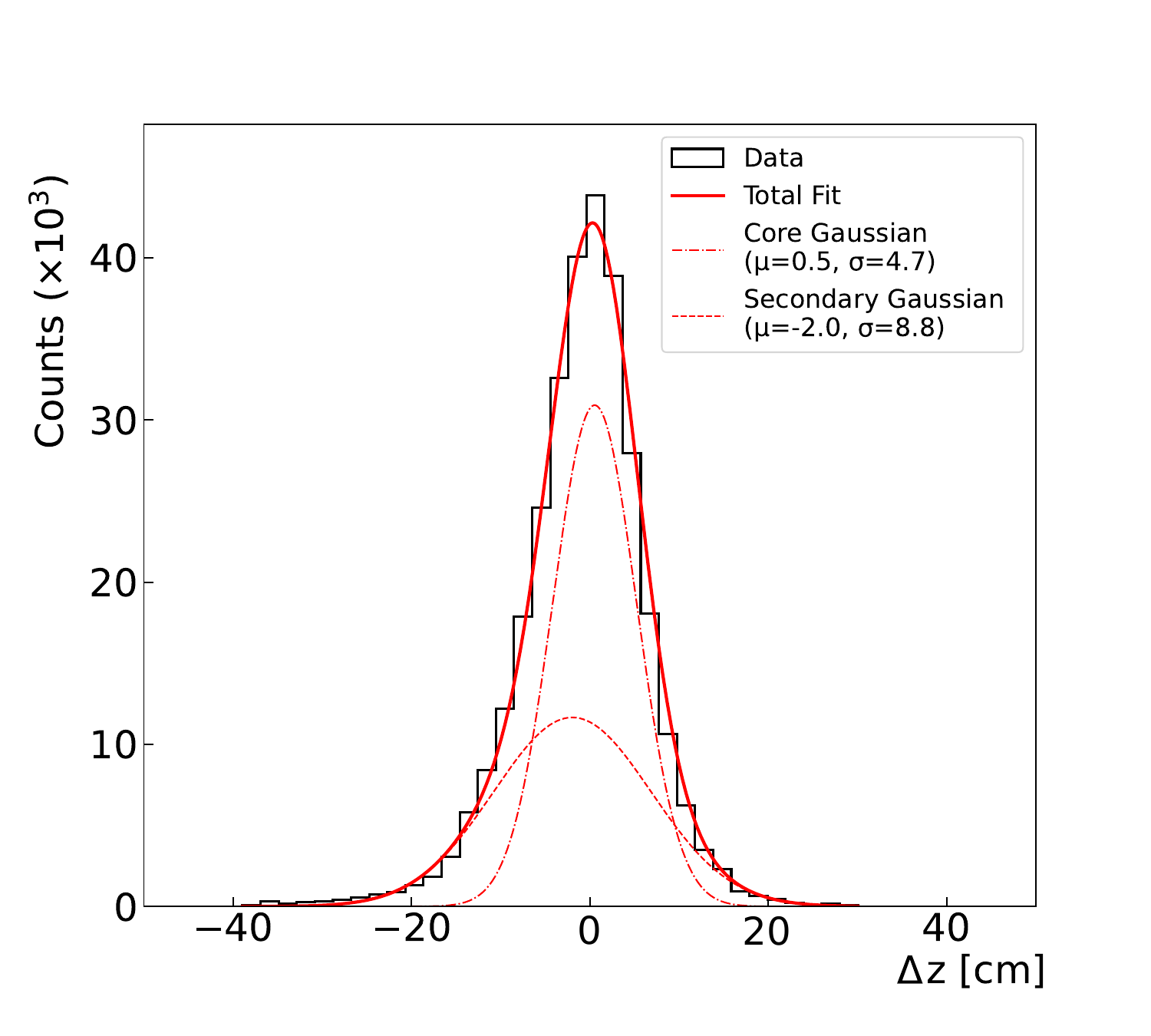}
\includegraphics[width=0.48\textwidth]
{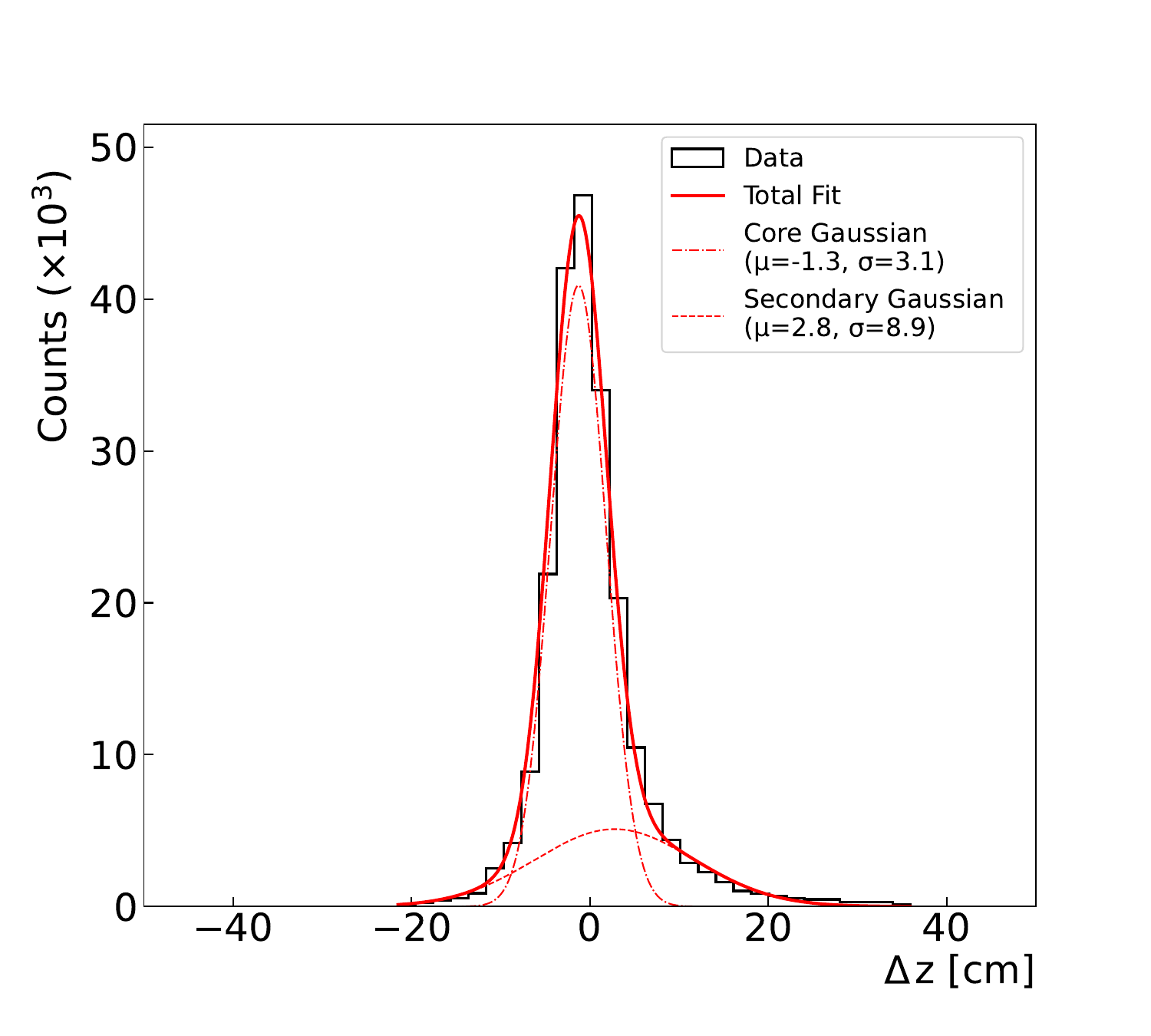}
\end{minipage} 
\caption{
  Difference of the $z$-positions between the reco track and the associated neuro track, in [cm], for Exp. 16 (initial FANN training) and Exp. 24 (presently used Pytorch training). The distributions are fitted with double Gaussians (see text).}
\label{fig:strackRes} 
\end{figure}

Despite the short periods of ``low'' background in 2021, it became clear that the neural trigger should learn to cope with increasing beam currents and backgrounds while the machine is struggling to increase the luminosity. For this, a new training with tracks from the high background runs of 2021 was launched, using PyTorch. The data for the training have been derived from the events which had passed any of the L1 trigger conditions (from the track and calorimeter trigger) and had by-passed the High Level Trigger selection. Although the background in these data was considerably higher, a significant improvement of the $z$-resolution  was observed resulting from the PyTorch training.
Figure~\ref{fig:strackRes} shows, for offline reconstructed tracks from Exp. 16 (FANN) and Exp. 24 (PyTorch), the difference in $\Delta z = z_{\rm neuro} - z_{\rm reco}$ between the reco tracks and the corresponding neuro tracks. 
Here, events with a single reco track within the CDC acceptance (transverse momentum $p_T > 250$ MeV, $-0.54 < \cos(\theta) < 0.82$) were required. The CDC conditions are necessary to make sure that tracks in the $r\phi$ plane can be found by the conventional 2D tracker, which is the required input to the preprocessing steps for the neuro tracks (see the hardware section above). In addition, a cut in the $z$-vertex for the reco tracks of $|z_{\rm reco}| < 1$ cm was imposed to make sure that only \epem collision events enter the sample. For comparison, the distributions are described by double-Gaussians. 
While the initial training with FANN showed a resolution in the core Gaussian of 4.7 cm, the new training yielded a significantly better value of 3.1 cm. The wide Gaussians (dashed lines in fig.~\ref{fig:strackRes}) with typical contributions of order 10\%, have widths around 8.8 cm, but with a much smaller contribution for the neuro tracks trained via PyTorch. Based on the improved resolution of the $z$-impact, a reduced, but still conservative, cut of $|z| < 15$ cm was chosen in the GDL for a neuro track to pass the track trigger.

\begin{figure}[!ht]
\begin{center}
\begin{minipage}[]{1.0\linewidth}
\centering
\includegraphics[width=0.49\textwidth]
{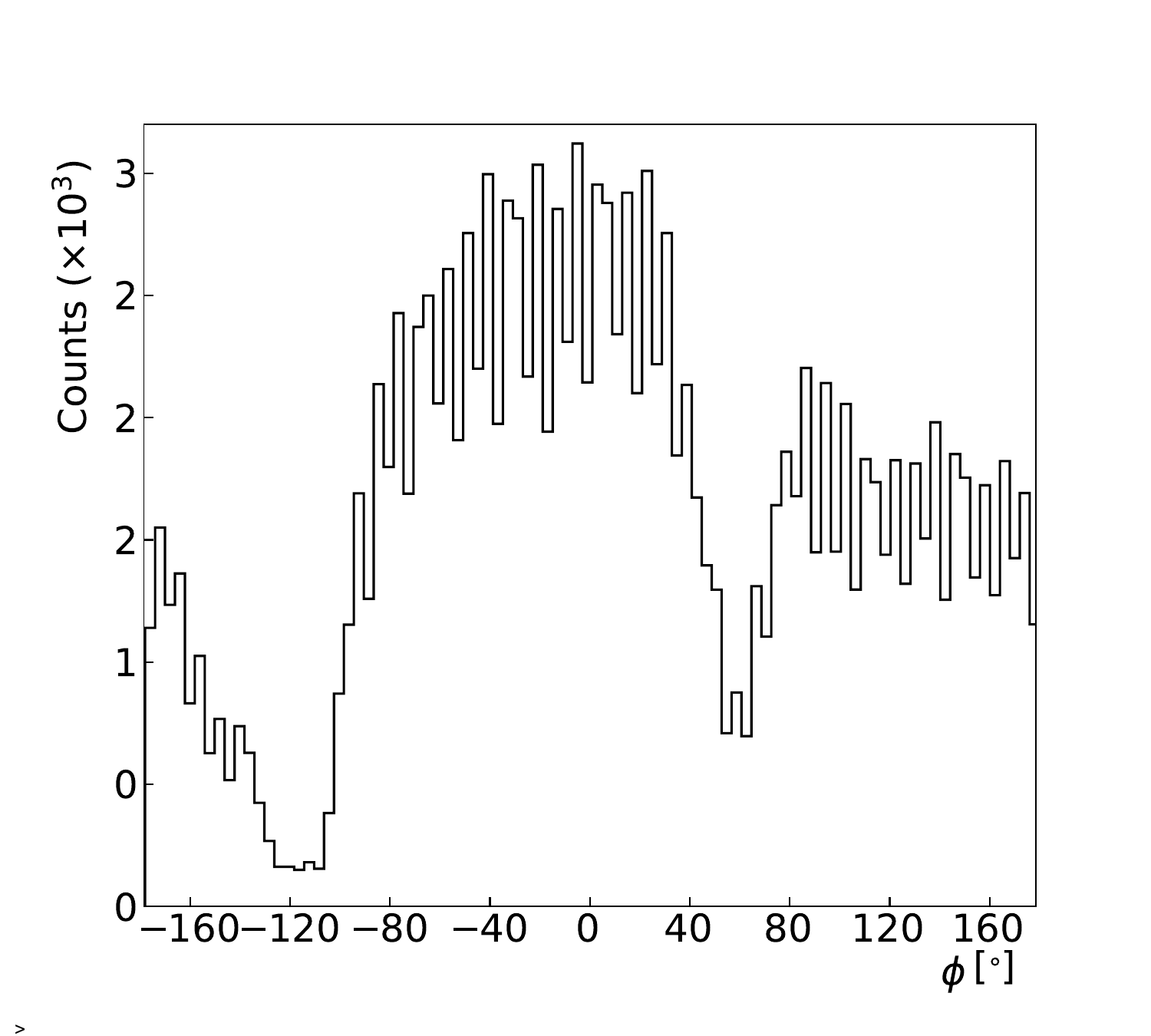}
\includegraphics[width=0.49\textwidth]
{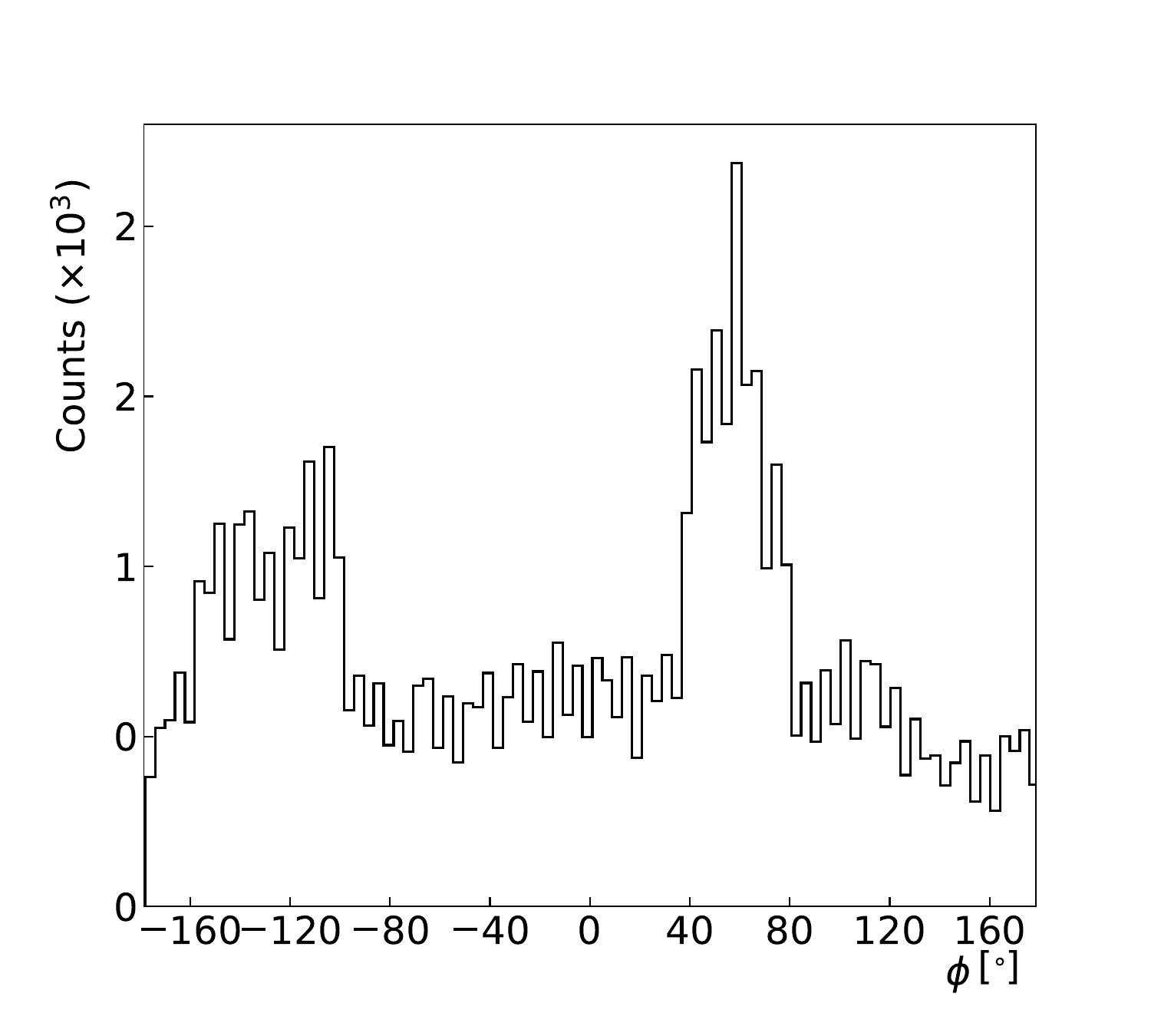}
\end{minipage}
\hfill
\begin{minipage}[]{1.0\linewidth}
\centering
\includegraphics[width=0.49\textwidth]
{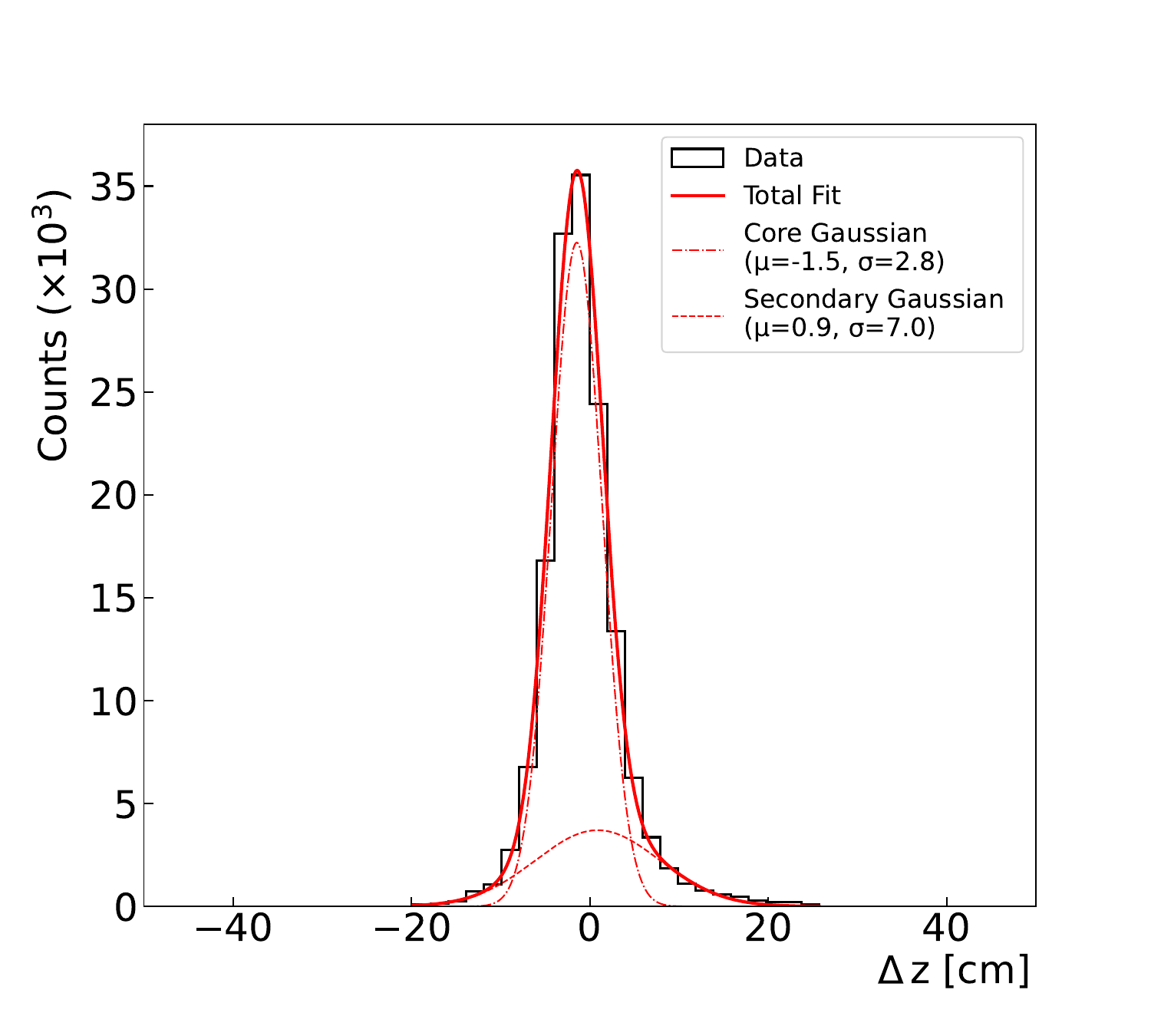}
\includegraphics[width=0.49\textwidth]
{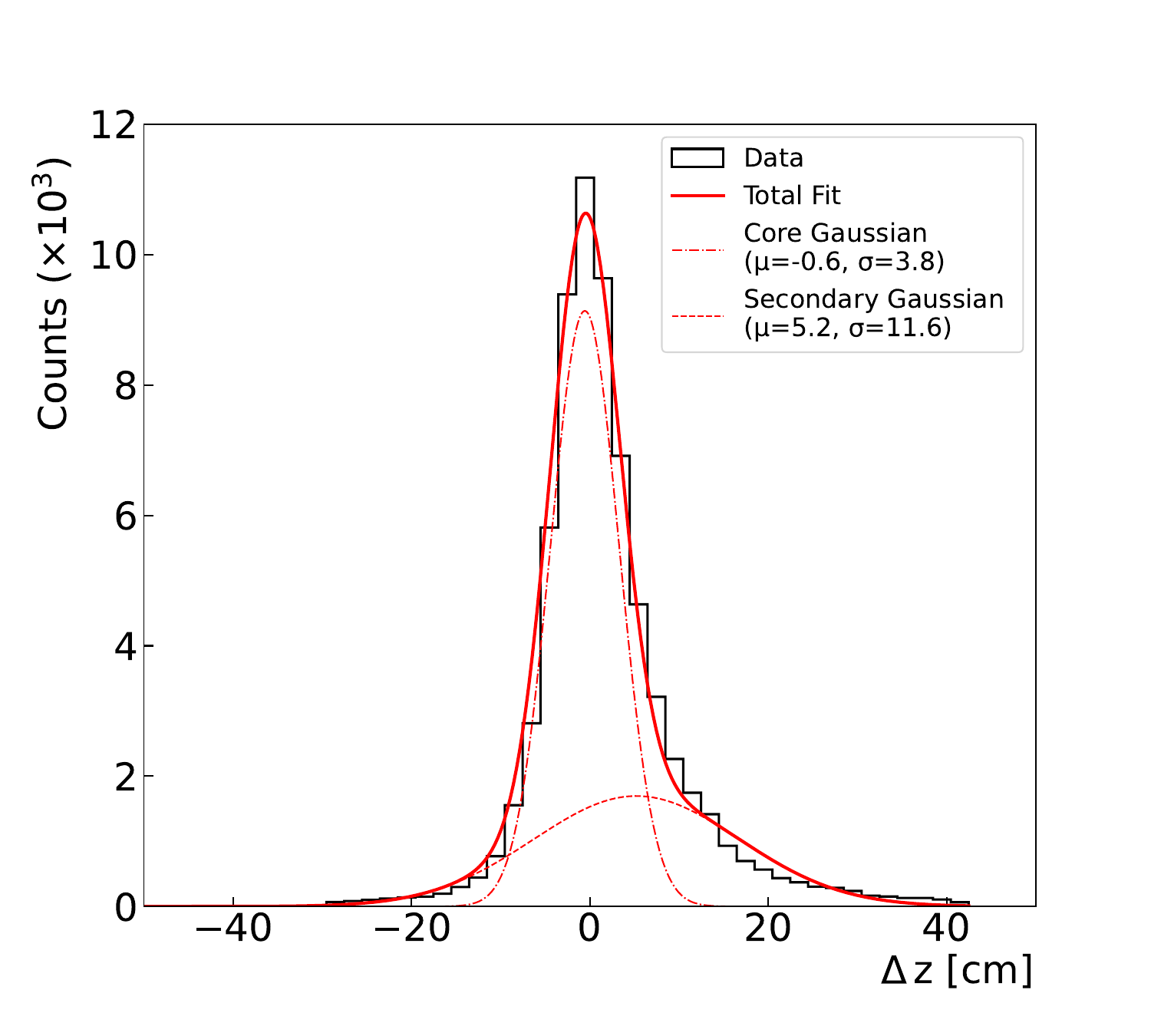}
\end{minipage}
\caption{Distributions of the azimuth angle $\phi$ and $\Delta z$ for the neuro tracks for different numbers stereo SLs. Left column: neuro tracks with all 4 SLs. right column: neuro tracks with one of the stereo SLs missing. Data from early 2022 running (Exp. 24). For about 70\% of the tracks signals from all stereo Sls were present in this experiment. }
\label{fig:phidist} 
\end{center}
\end{figure}

As explained earlier, the system of neural networks delivering $z$ and $\theta$  for the tracks in an event consists of a total of five expert networks. While the expert net ``0`` is used for ``clean'' neuro tracks, where all 4 SLs are present, the other experts (``1 - 4'') are chosen depending on which one of the stereo layers is missing. The cooperation of the expert networks is illustrated in fig.~\ref{fig:phidist}. On the left column of the figure the distributions of the azimuth angle $\phi$ (top) and $\Delta z$ (bottom) for the expert net 0 are shown, on the right column the corresponding distributions for the experts 1-4. The data are taken from Exp. 24, where only the neuro tracks associated with reconstructed tracks from the origin ($|z| < 1$ cm) are selected. One observes two regions of inefficiencies in the $\phi$ distribution for the expert 0. The holes are caused by local inefficiencies of the stereo drift wires during the data taking period of Exp. 24. These areas of inefficiencies are nicely ``filled'' by the experts 1-4 so that the combined $\phi$ distribution shows the expected uniform shape, slightly peaking at zero degrees due to the non-zero crossing angle of the electron positron beams.    
        
The bottom row of fig.~\ref{fig:phidist} shows the $z$-resolution for clean tracks (left) compared to their complement on the right. The fraction of the clean tracks is typically around 70-80\%. The $z$-resolutions are fitted by  double-Gaussian functions. The resolutions of the core Gaussians are 2.8 cm for the clean tracks, and 3.8 cm for all the other tracks. This increase is understandable in view of the fact that fewer stereo SLs are available for the $z$-estimate. In particular the tracks from the ``Expert 4'' network have worse resolutions where the innermost  stereo layer. i.e. the one closest to the track origin, is missing. A typical share between the expert networks, reflecting the inefficient regions in the CDC front end electronics before LS1, is 86\% for Expert 0 and 2\% each for the ``inner '' Expert networks 2, 3, and 4. When the outer stereo TS is missing (Expert 1, typically 5-8 \%), then the reason is a combination of CDC inefficiencies and shallow polar angles of the tracks, missing the full angular acceptance of the CDC.    

\begin{figure}[!htb]
\begin{center}
    \begin{tikzpicture}
        \node[anchor=south west,inner sep=0] at (0,0) {\includegraphics[width=0.8\textwidth]{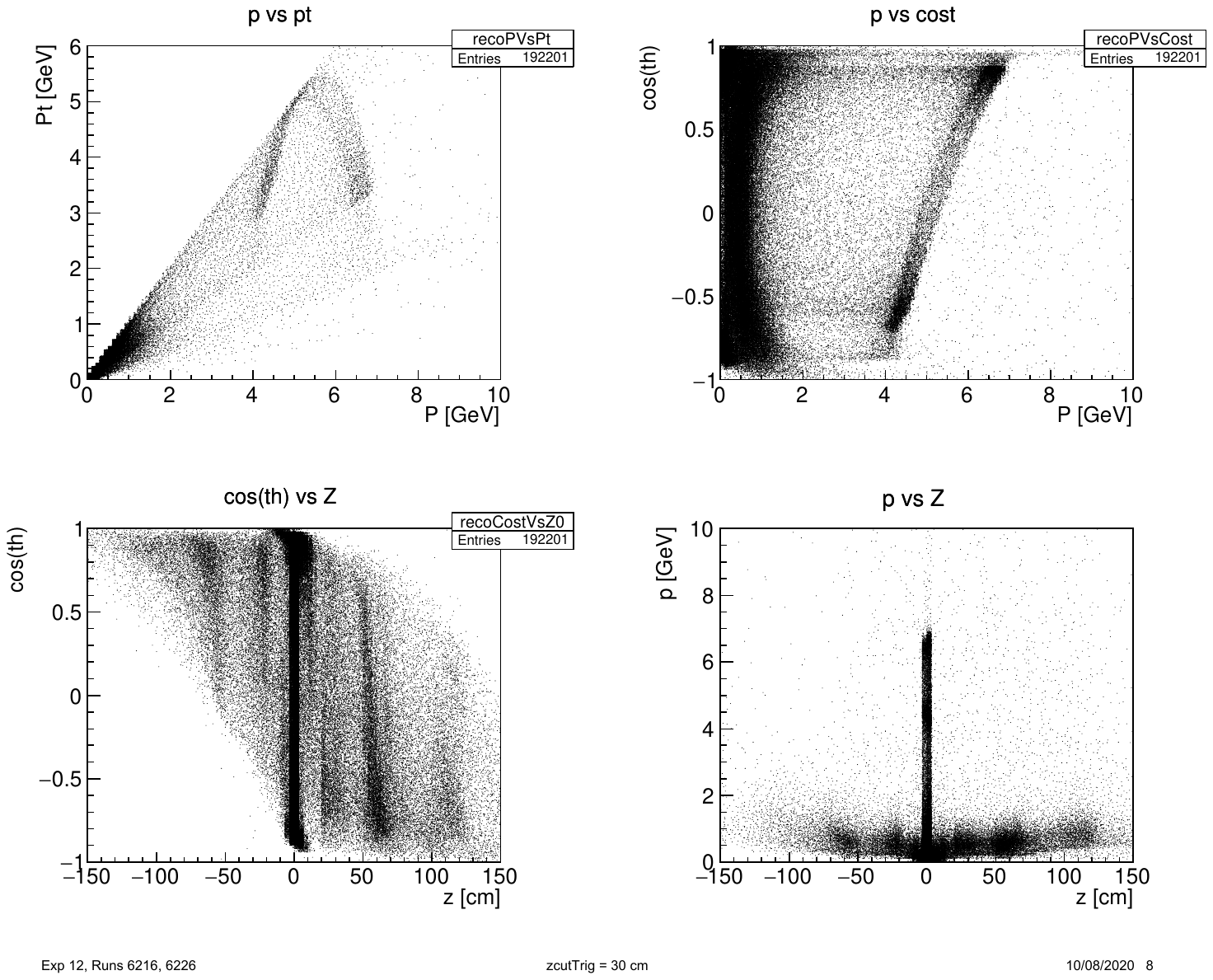}};

        \draw[red, thick] (1.65,1.9) -- (11,1.9); 
    \end{tikzpicture}
\caption{ 
  Distribution of the track momentum versus the origin in $z$ for all tracks fully reconstructed in the event. A cut at 0.7 GeV (red line) gets rid of QED and also of most of the tracks from outside the IP (data from Exp. 16). }
\label{fig:pvsz} 
\end{center}
\end{figure}

\paragraph{\bf {Single Track Trigger (STT)}:}
In addition to the requirement of a neuro track for all standard $\ge 2$-track triggers, it was possible to launch a minimum bias single track trigger, requiring $|z| < 15$ cm, supplemented by an additional minimal requirement for the track momentum of $p > 0.7$ GeV. 
The latter cut is necessary to suppress a large ``background'' coming from IP: 
Figure~\ref{fig:pvsz} shows the track momentum $p$ versus the track origin $z$ in all events of Exp. 16 after full track reconstruction. The large peak around $z=0$ at low momenta comes mainly from the QED reaction \epem $\rightarrow$ \epem \epem, where the secondary electrons and positrons peak at extremely low energies. The momentum cut largely reduces the QED background, but also strongly reduces the off-IP background generated by spallation protons, with typical momenta of 500 MeV. 

At the trigger level, the momentum of the neuro track is calculated from the curvature $\omega$ of the input 2D track, the known solenoid field $B$ of \belleii, and the polar scattering angle $\theta$, estimated by the network (second output of the networks, see fig.~\ref{fig:NetArchitecture}): 
$$
p[\rm{GeV}] = {{1} \over {\omega[1/\rm{m}]\;\sin(\theta)} \;0.3 \; \it{B} [\rm{T}]} .
$$

\begin{figure}[!ht]
\begin{center}
\begin{minipage}[]{1.0\linewidth}
\centering
\includegraphics[width=0.49\textwidth]
{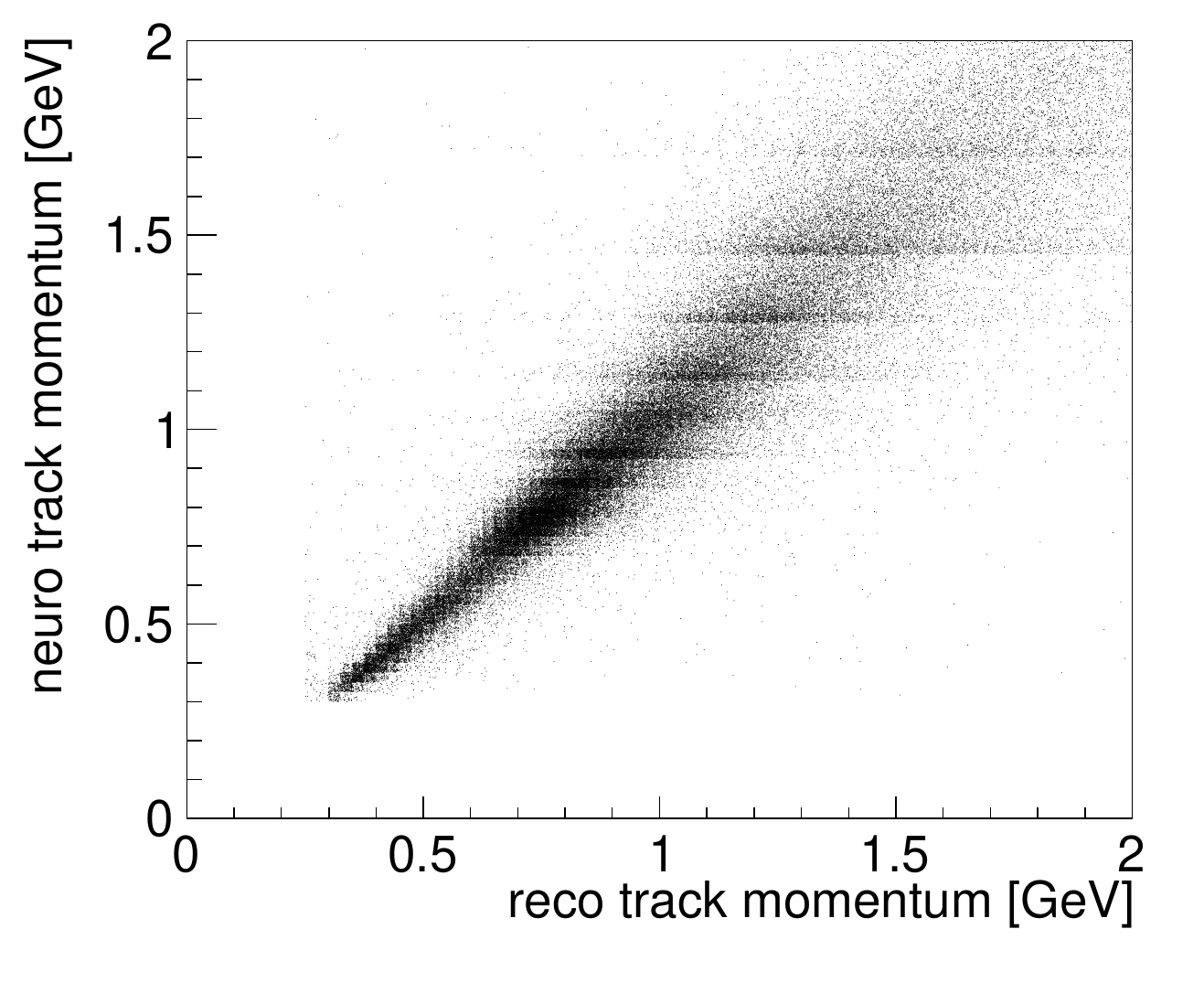}
\includegraphics[width=0.49\textwidth]
{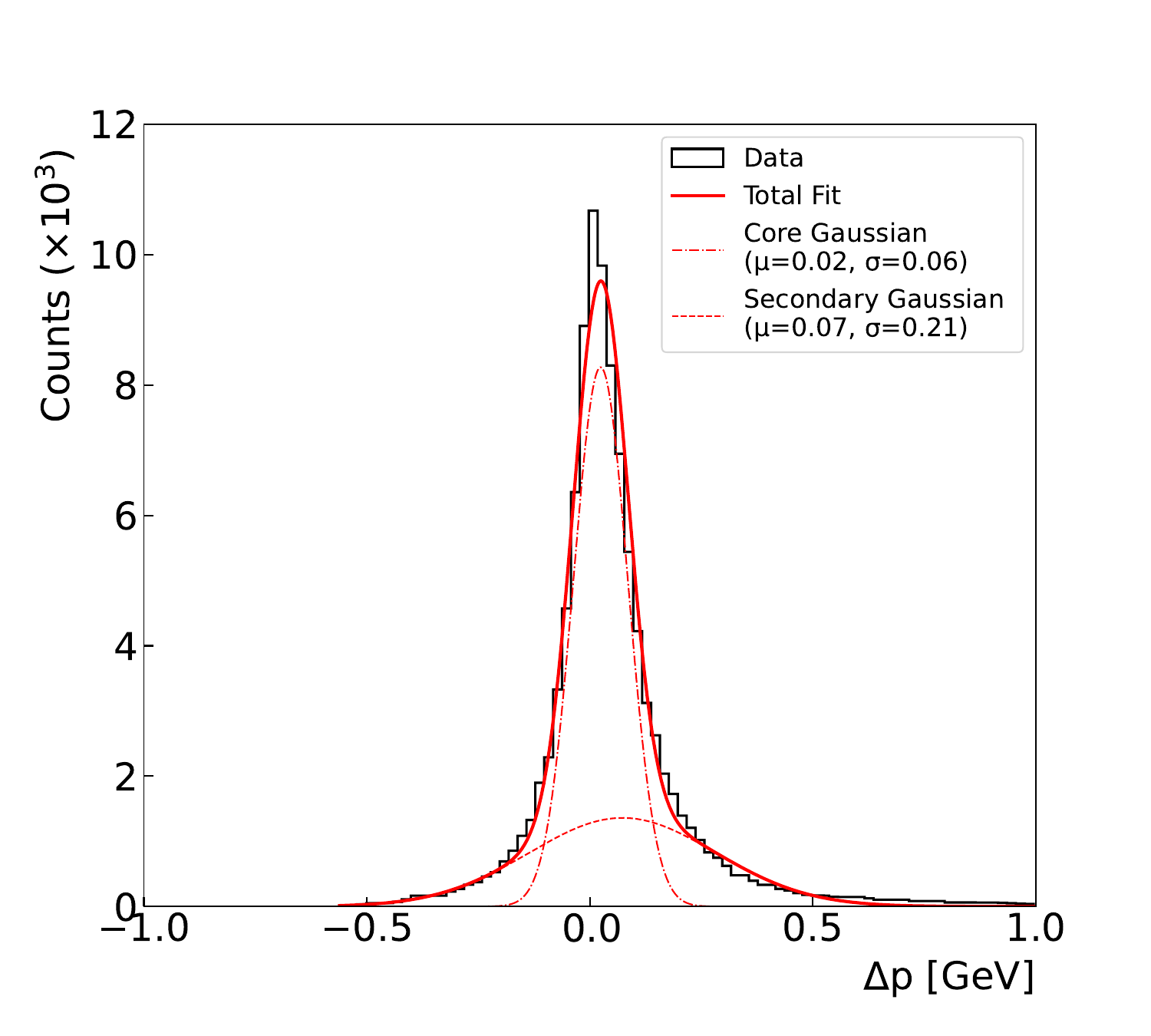}
\caption{ 
  Left: Correlation of the true track momentum from reconstructed tracks and the momentum estimated for the neural tracks. Right: Difference of track momenta (see text).}
\label{fig:presolution} 
\end{minipage}
\end{center}
\end{figure}

Based on this expression the momentum of a neuro track is determined in the GRL using a look-up table for the trigonometric function 
$\sin(\theta)$. 
The momentum resolution of the neuro tracks is shown fig.~\ref{fig:presolution}. On the left side the correlation between the true momentum (after full offline reconstruction) with the result for the corresponding neuro track is shown, based on the above expression. The horizontal stripes are due to the integerized hardware representation of the 2D track candidates.  
The tracks selected are from events triggered by the STT towards the end of the running before LS1, which was plagued by very high backgrounds (``Exp. 26''). 
On the right side of the figure the difference between the true and neural momentum is displayed. A fit to the distribution with a double Gaussian yields a standard deviation of about 60 MeV for the core Gaussian. As can be seen from the correlation plot, the core Gaussian mainly describes the resolution of momenta below 1 GeV.    

The STT was launched as a physics trigger (unprescaled) for low multiplicity events since the beginning of the year 2021. Its contribution to the total trigger rate was observed at the level of roughly 20\%, which is very reasonable for a minimum bias trigger. Since only a single track is required, further tracks have no limitation set by the acceptance of the CDC. Therefore also 2-track events are triggered where the second track is only going through a few planes of the CDC and the offline reconstruction is mainly done by the vertex detectors. 

\begin{figure}[!ht]
\begin{center}
\includegraphics[width=1.0\textwidth]
{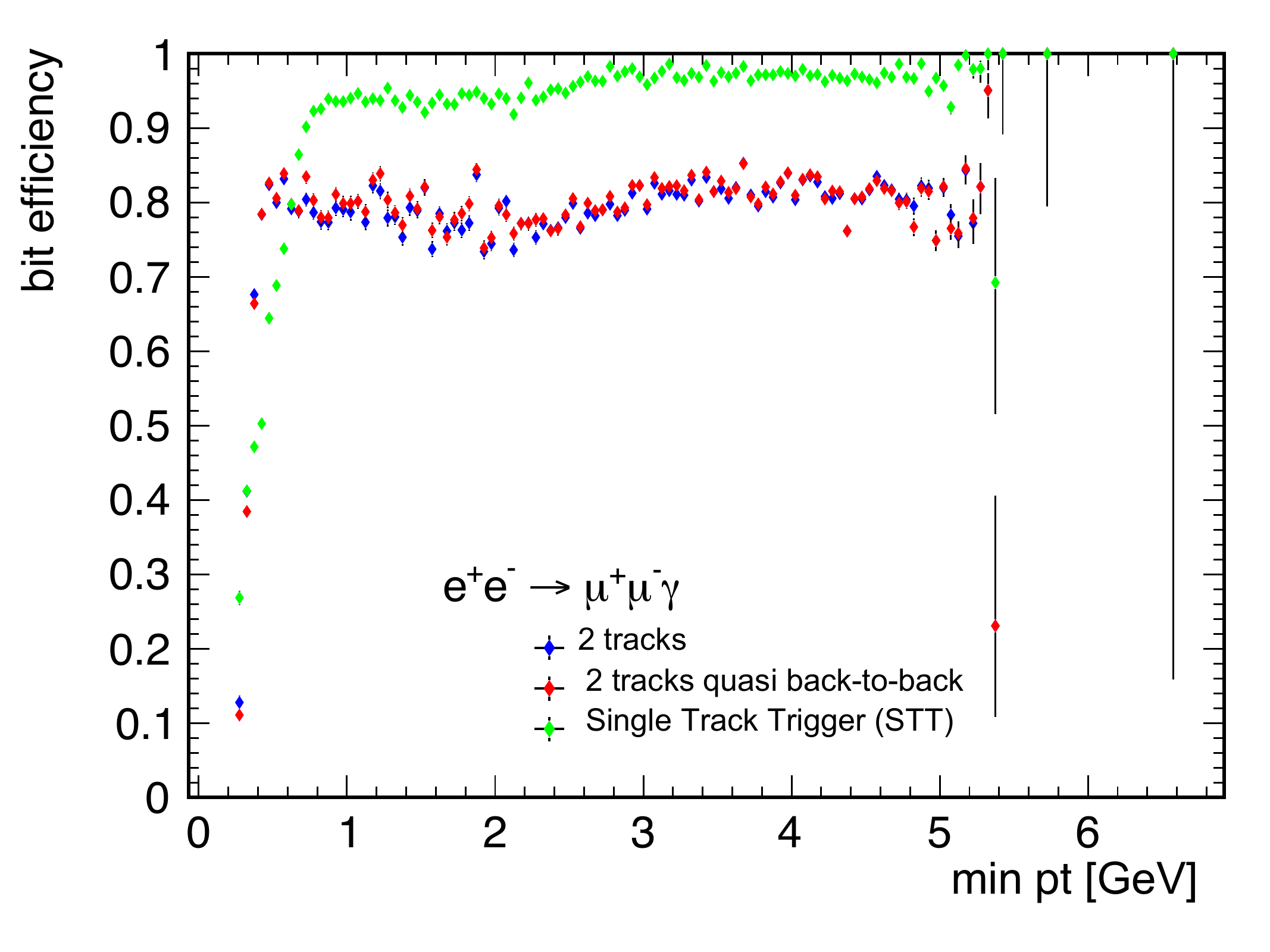}
\caption{Efficiency of STT in comparison to the two-track triggers for the reaction \epem $\rightarrow \mu^+ \mu^- (\gamma)$, as function of the smaller transverse momentum of the two tracks.}
\label{fig:stteff} 
\end{center}
\end{figure}

A comparison of the largely improved efficiency of the STT relative to the other multi-track triggers (still requiring a neural track, see above) is shown in fig.~\ref{fig:stteff} for the reaction \epem $\rightarrow \mu^+ \mu^- (\gamma)$. The plot shows the efficiency as a function of the track with the smaller transverse momentum. The observed improvement of the efficiency for momenta larger than 0.7 GeV is evident.   
 
Belle\,II has successfully taken physics data with the neural L1 trigger, running now at world record luminosity exceeding $4 \times 10^{34}$ cm$^{-2}$s$^{-1}$. SuperKEKB aims to eventually provide instantaneous luminosities beyond ${\cal L} = 6\times 10^{35}\,\mathrm{cm}^{-2}\,\mathrm{s}^{-1}$. The pure physics rate at this super-high luminosity is expected to be around 10\,kHz, with a limitation of the maximal data logging rate of 30\,kHz. This means that the L1 triggers have to be selective enough to accept a background over signal fraction of only 2 to 1, a condition for both the calorimeter and track triggers. This will be a very demanding task which is by no means realized at present, and will depend strongly on the operating conditions of SuperKEKB at that time. 

Towards the end of the data taking period, before the Long Shutdown LS1 in June 2022, the machine conditions for these runs (Exp. 26) were characterized by extremely large background in the CDC, which led to increased rates for the \ztrig, dominated by the STT: Instead of the usual 20\% share in the total trigger budget, this share has increased to about 50\%. On the other hand, the efficiencies for vertex tracks did not change within their uncertainties: Although the networks have been retrained using data with the increased background, the training samples are largely dominated by vertex tracks (see fig.\ref{fig:allzplots}, lowest row). 

\begin{figure}[!ht]
\begin{center}
\includegraphics[width=1.0\textwidth]
{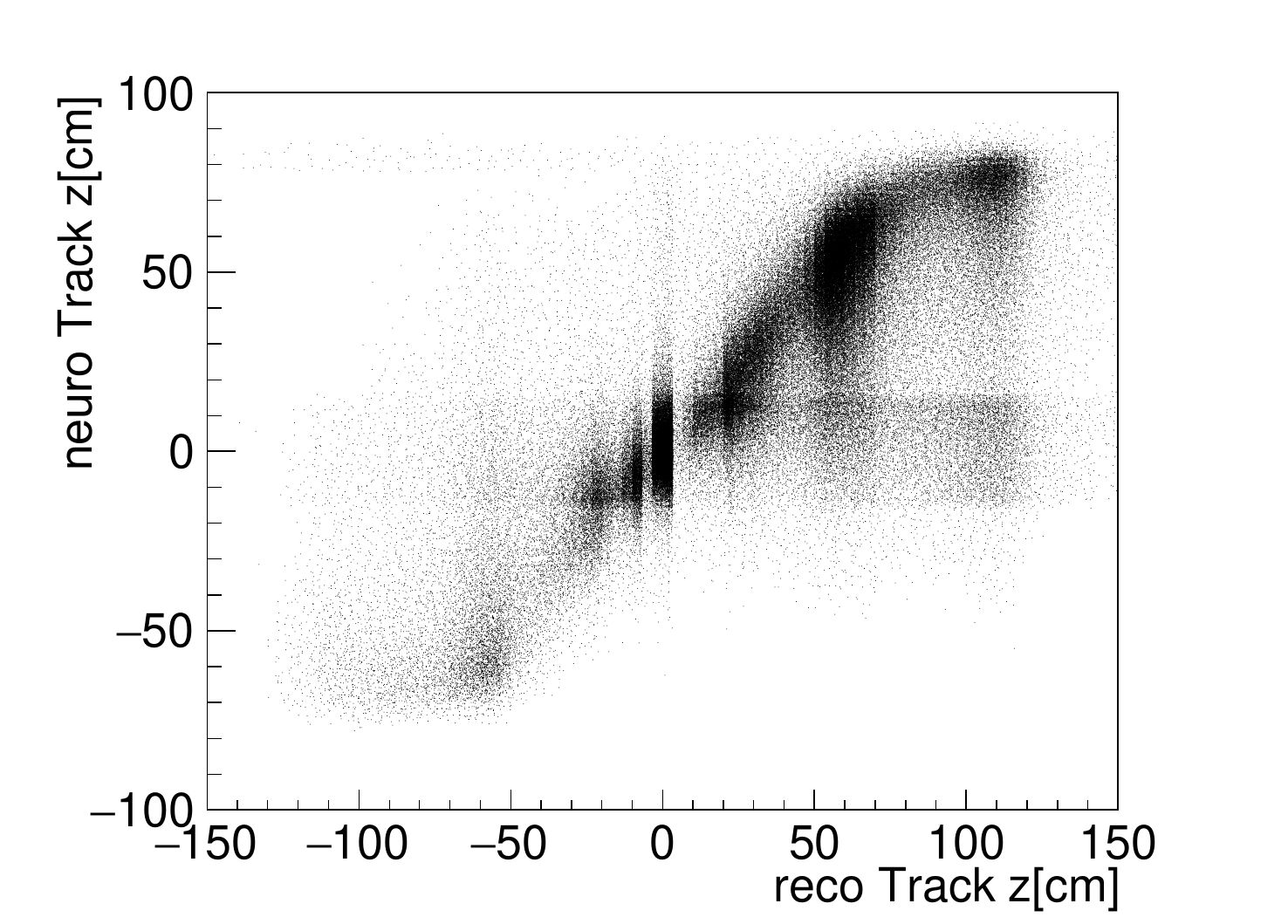}
\caption{Correlation of the $z$-impact between the fully reconstructed tracks and the corresponding neuro tracks.  The horizontal band between $\pm$ 15 cm marks the acceptance region of the single track trigger. The ``feed-down'' effect into this band, mainly from the $z$ intervals $\pm$ 20 cm and $\pm$ 50 cm, is clearly visible. The data were taken in a short run where the \ztrig was not enabled.}
\label{fig:zzcorr} 
\end{center}
\end{figure}

Given the stable $z$-resolution of the \ztrig for tracks from IP under high background conditions, we still need to explain the increase of the trigger rate. To this end we present in fig.~\ref{fig:zzcorr} the correlation between the $z$-values for the reco tracks with those for the corresponding neuro tracks. 
In this $z$-correlation plot one observes ``feed-down'' of the real tracks with $z$-values larger than about 20 cm into the $z$-acceptance interval of the neuro tracks ($|z| < 15$ cm). 
These additional neuro tracks are clearly visible in the horizontal acceptance band for neuro tracks at $\pm$ 15 cm. They come from background, but will set the \ztrig and therefore increase the L1 track trigger rate. 
In addition, we also observe an increasing number of fake neuro tracks with $z$ values inside the acceptance band, which do, however,  not correspond to any real reco track. 
Both effects, feed-down and fake tracks are generated mainly by an increasing rate of 2D input track candidates, which are formed by or contaminated with random background hits. 
These 2D tracks have a fair chance to be combined with stereo track segments originating from background sources, producing valid neuro tracks within the required $z$ acceptance. 
The observed sensitivity of the \ztrig to increased background rates is in part understandable since so far a relatively small fraction of tracks at $z$-values further away from IP are available for the re-training of the networks (see, e.g., the last row in fig.~\ref{fig:allzplots}). 

Our ongoing studies to improve the $z$-resolution for the entire $z$ region ($\pm 100$ cm) with the aim to significantly reduce the feed-down and fake tracks effects, while keeping the physics efficiency high, are the subject of the next section.

\section{Ongoing Developments}
\label{sec:future}

We envisage several ways to stabilize the STT and the multi-track \ztrig for future running (clearly, we exclude the possibility to simply pre-scale the STT and lose physics). Since new and more powerful custom-made trigger boards (``UT4'', equipped with Virtex UltraScale 7 XCVU080/160 FPGAs) have become available to us  recently for the \ztrig, more resources are now available to overcome the limitations of the presently installed UT3 trigger boards. This means that the neural network architecture of the \ztrig, limited at the moment to one hidden layer with 81 nodes only, can now be extended to a deep-learning network model, having typically three to four hidden layers with $\cal{O}$(100) nodes each. Furthermore, the track segment finders (aTSF and sTFS, see fig.~\ref{fig:hw:cdcarch} above) will also provide information on all other wires within the TSs in addition to the priority wire: This additional information consists of binary information of the charge measured on the wires as well as the \drift, although with somewhat reduced precision in the \drifts (32 ns instead of 2 ns). Adding the information from the additional wires within a TS, we expect better performance due to the additional constraints. This entails, of course, to widen, possibly substantially, the input layer.

\begin{figure}[!ht]
\begin{center}
\begin{minipage}[]{1.0\linewidth}
\centering
\includegraphics[width=0.49\textwidth]
{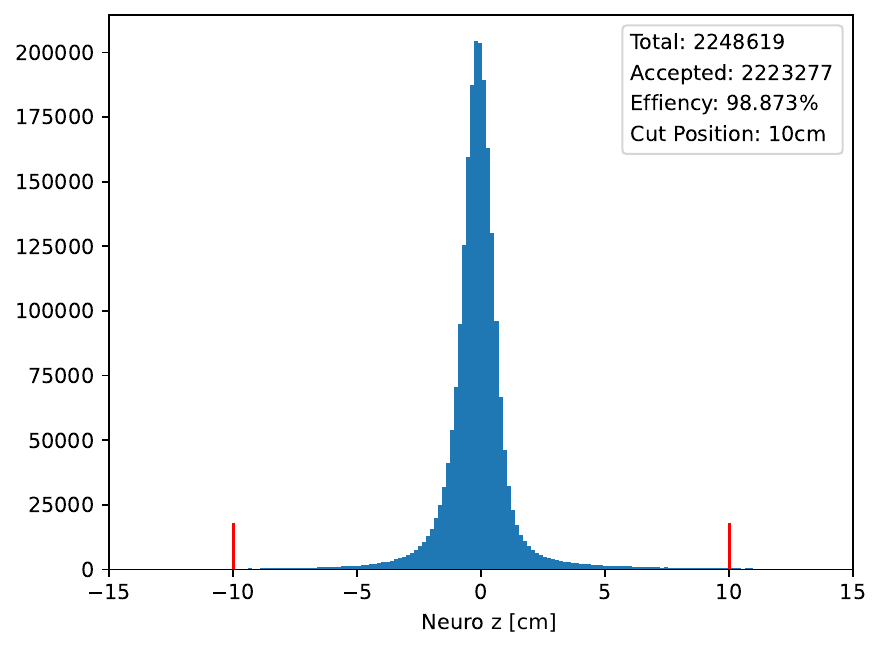}
\includegraphics[width=0.49\textwidth]
{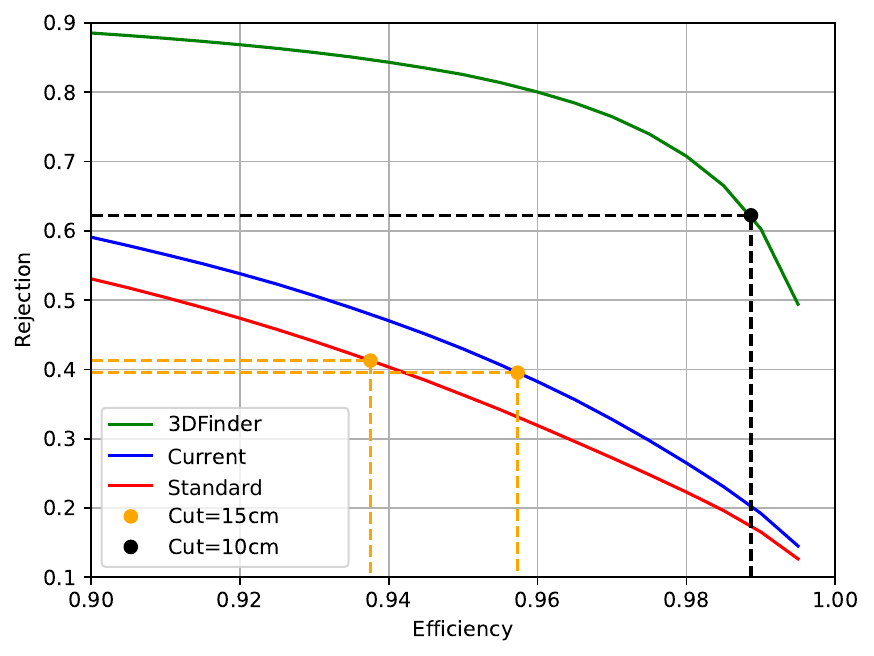}
\caption{ 
Expected performance of a deep-learning network architecture with 3D track candidates as input plus topological information from all wires in the associated track segments (``3DFinder''). Left: $z$ estimate of the 3DFinder network for IP tracks, right: rejection of background vs. efficiency for IP tracks, green curve), compared to the present networks in blue and red (see text).}
\label{fig:upgrade} 
\end{minipage}
\end{center}
\end{figure}

Most importantly, the input track candidates for the networks need to be made robust against background. Presently the track finding is carried out by the 2DF module (see Fig.~\ref{fig:hw:cdcarch}), using Hough transforms in the space spanned by the azimuth angles $\phi$ and the track curvature $1/r$. For a valid 2D track candidate at least 4 out of the 5 axial layers are required. To reduce the chance of track candidates formed by background TSs, we have investigated~\cite{SSkambraksDiss} to use in addition the priority wires of the stereo TSs. In this scheme the 2D Hough plane is enlarged to a three-dimensional Hough space, adding as a third dimension the polar track angle $\theta$. The enlarged Hough space has the advantage that now all nine SLs (instead of only the 5 axial SLs at present) can be used for track finding, making random noise much less likely. As a further advantage of the 3D Hough space the track candidates are forced to come from IP, thus naturally suppressing tracks from outside. 
We found out that even under very severe background conditions, extrapolated to the design luminosity, the 3D Hough model finds the correct tracks with high efficiency and, at the same time, strongly reduces the feed-down effect and fake tracks. 

In our studies we use all wires in the TSs, summing up to a total of 126 inputs from all nine superlayers (compared to 27 at present). For the networks we tried several deep-learning architectures, implementing 3 and 4 hidden layers with 60 to 80 nodes each (more hidden layers are not possible due to latency requirements in the UT4). We also tried additional expert networks with deep layers. But it turned out that they did not improve the results of a single deep-learning structure. The training sample was again taken from the high background running in Exp. 26. Using the 3D Hough track candidates, the additional inputs from all wire cells within the TSs, and a deep-learning architecture for the neural networks, we could demonstrate that the $z$-resolution of the vertex tracks can be further improved, and the background (feed-down and fake tracks) can indeed be substantially suppressed. 

The results of these studies are shown in Fig.~\ref{fig:upgrade}, where we compare our new studies (labelled ``3DFinder'') with the performance of the presently implemented set of networks (labelled ``Current'', ``Standard''): On the left side the $z$ resolution is shown for a 4-hidden layer, fully connected network with 60 nodes each (``3DFinder''). One sees that the $z$ cut can easily be reduced to 5 cm. On the right side the expected performance of ``3DFinder'' (green line) is compared with the single hidden-layer networks running before the Long Shutdown 1 (``Standard'', red curve). The blue line is the result after re-training the ``Standard'' networks with the same high background data (Exp. 26) as ``3DFinder''. The curves show the efficiency versus background rejection on an independent test sample from Exp. 26. The curves are generated by varying the $z$ cut, both cuts ($|z| < 15$ cm and $|z| < 10$ cm) are indicated as points on the curves. 

There are also studies going on to implement a neural trigger, working in parallel to the \ztrig, to identify track pairs with a strongly displaced vertex (``Displaced Vertex Trigger'' DVT). Such event signatures are expected, for example, in models that predict inelastic dark matter production in association with a dark Higgs boson~\cite{idmmodels}. The method of choice for the track finding at L1, given the wire preprocessing blocks up to and including the track segment finders (see fig.~\ref{fig:hw:cdcarch}), is the 2D Hough transform again. Here, however, a vertex hypothesis is necessary to create the intersecting curves in the Hough space. For the \epem collision events the case is clear: the vertex is the IP, well approximated by $(x=0,y=0)$. For the displaced vertices, however, the origin is unknown. The method of choice is to put a grid of possible vertex locations (we call them ``macro cells'') over the transverse cross section of the CDC. A typical number of the finite-size macro cells is of order 100, where the Hough transforms have to be executed all in parallel with each of these macro cells as vertex hypothesis. But then it turns out that simple peek finding in the Hough space will not be selective enough to pin down the correct macro cell. The solution found here is to employ neural networks to analyze the Hough cluster shape, similar to the way described above for the 3D track model. In the new UT4, the parallel execution of all Hough transforms seems possible within the latency attributed to the L1 track trigger.    

\section{Summary, Outlook and Conclusions}
\label{sec:conclusions}

For the Belle~II experiment, running at the SuperKEKB asymmetric-energy electron-positron collider, the first fully operating global L1 track trigger based on neural networks has been realized. 
The neural trigger uses the information from the axial and stereo wires of the Central Drift Chamber (CDC). 
The input to the networks are the track candidates provided by the original \belleii track trigger, which are found by means of Hough transforms in the plane transverse to the beams. 
This 2D trigger is unable to reject the dominating track rate generated by beam collisions with the beam pipe and structures of the beam focusing system. 
Adding the stereo wires of the CDC, a set of single hidden layer networks provide as outputs the vertices of the 2D track candidates along the beam ($z$)-direction and their polar emission angles $\theta$.
With this information the neural trigger identifies the events coming from outside of the electron-positron interaction point ($z=0$). 
All track triggers with two or more 2D candidates  require at least one neural track with the condition $|z| < 15$ cm. 

Using the polar angle $\theta$ it is even possible to deploy an unprescaled, minimum-bias single track trigger (STT), which requires a mild momentum cut of 0.7 GeV in addition to the acceptance cut of $|z| < 15$ cm. The STT is particularly effective in selecting events with low charged multiplicity.  

Although the STT has boosted the track trigger efficiency, it showed some sensitivity to the strongly increasing backgrounds with the machine's program to raise the luminosity towards design values. 
While the performance of the STT for physics triggers was not affected by the high backgrounds, demonstrating the robustness of the neural approach, it showed an over-proportional increase in trigger rate, generated mainly by the background-prone 2D track finding algorithm.
In order to provide stable operation for the STT in the future, a comprehensive upgrade program is in progress, replacing the traditional 2D track finding by a novel three-dimensional Hough space analysis, and an extension of the network structure to a deep-learning architecture with additional wire inputs. We have shown the expected increase of performance of this new scheme. 

Since the neural triggers so far are optimized for tracks coming from the interaction point, they will miss event signatures with tracks coming from a strongly displaced vertex. Such event signatures are expected in various extensions of the standard model that predict long-lived particles. We shortly sketched our plans to add neural algorithms to the \belleii L1 track trigger system for such anomalous event types.

The global L1 neural network hardware track trigger developed for \belleii and reported in this paper represents, in our opnion, a significant step towards fast and intelligent data selection and processing methodologies in high-energy physics experiments. Our approach to successfully integrate machine learning techniques via flexible FPGA hardware in a running experiment points towards a possible paradigm shift, embracing smarter and faster solutions for real-time event selection. It also sets a precedent for future experiments where efficient and accurate data collection as well as extraction of novel features are essential.



\section*{Acknowledgments} 
The authors would like
to thank the members of the Belle II trigger group for their valuable contributions, 
discussions and suggestions.  This work is funded by the German ministry of science and education BMBF (Verbundprojekt 05H2021, ErUM-FSP T09). 
G.\,Inguglia and M.\, Bertemes acknowledge funding from the ERC - European Research Council (StG nr. 947006) and from the FWF - Der Wissenschaftsfonds (project nr. J 4625), respectively.

\end{document}